\documentclass[letterpaper,11pt]{article}

\usepackage{amssymb}
\usepackage{amsmath}
\usepackage{amsthm}\usepackage{graphicx}
\usepackage{authblk}
\usepackage{xcolor} 

\usepackage{float}

\usepackage{booktabs,siunitx}
\usepackage{multirow}

\usepackage{fullpage}
\usepackage{bm}

\usepackage[utf8x]{inputenc}

\usepackage{color}
\usepackage{algpseudocode, algorithm, algorithmicx}
\usepackage[normalem]{ulem}

\newcommand{\jh}[1]{\textcolor{black}{#1}}

\newcommand\amsclass[1]%
  {\hspace{9mm}
   \textbf{MSC codes.  }#1
}

\usepackage{romannum}

\title{A derivative-free framework for capturing macroscopic behavior of incompressible turbulent flows}
\author[1]{Jihun Han\thanks{jhan25@albany.edu}}
\author[2]{Yoonsang Lee\thanks{yoonsang.lee@dartmouth.edu}}

\affil[1]{Department of Mathematics and Statistics, University at Albany - State University of New York, Albany, NY 12222}
\affil[2]{Department of Mathematics, Dartmouth College, Hanover, NH 03755}

\date{}

\begin{document}
\pagenumbering{arabic}

\maketitle
\begin{abstract}
The derivative-free loss method (DFLM) \jh{is a mesh-free neural network approach for solving} partial differential equations by simulating stochastic walkers rather than computing derivatives directly. Although DFLM has previously been applied to the Navier--Stokes equations, extending it to turbulent flows reveals two limitations of its simplest implementation\jh{. The} walkers do not account for the swirling, directionally stretching motion of the flow, and randomly sampling the walkers used to evaluate the target introduces noise into the learned solution. We address both limitations with an analytic approximation of the walkers' local motion that captures directional stretching and eliminates this sampling noise, while remaining computationally efficient. We study the resulting method as a non-intrusive multiscale solver capable of learning macroscopic flow behavior without resolving every scale on a fine grid or explicitly coupling coarse- and fine-scale models. For two turbulent two-dimensional flows, the proposed method reproduces the macroscopic energy spectrum of fully resolved reference simulations more accurately than the standard method, particularly where directional stretching is strongest, confirming DFLM's efficacy as a multiscale solver for turbulent fluid systems.

\end{abstract}



\section{Introduction}
Fluid systems in many engineering and geophysics applications, which are often described by the Navier-Stokes equation, are well known for a wide range of scales, hindering accurate computer simulation of all relevant scale components \cite{pope2000turbulent}. The energy backscatter from the small to large scales \cite{piomelli1991backscatter} must be accounted for to capture the correct large scale dynamics, while truncating small scale components leads to inaccurate macroscopic large scale dynamics. Instead of resolving all relevant scales, analytic or computational modeling methods that aim for capturing large macroscopic behaviors of the fluid systems have attracted many researchers, leading to various methods.

Large Eddy Simulation \cite{LES} closes the macroscopic resolved system through a modeling of the effective behavior of the unresolved scales, either through a nonlinear deterministic model \cite{bardina1980improved,meneveau2000scale,germano1991dynamic} or a stochastic model \cite{leith1990stochastic,marstorp2007stochastic, mason1992stochastic}.
In geophysical systems, the effect of the unresolved scale is sometimes modeled as stochastic forcing terms \cite{berner2009spectral, jansen2014parameterizing, mana2014toward}.

With increasing computer power, more nonintrusive computational approaches that bypass the modeling of the effective subgrid scale dynamics have been proposed. These methods instead simulate the subgrid scales directly, but only in local spatiotemporal domains that do not cover the whole domain, using the resulting local statistics to estimate the parameters needed to close the macroscale dynamics. The Heterogeneous Multiscale Method type approaches \cite{HMM,sandham2017surface} are examples of such methods. Superparameterization (SP) \cite{grabowski1999crcp,GroomsMajda2013SSP,SSPparam} follows the same strategy, embedding local high-resolution simulations within a coarse macroscale model to estimate the unresolved dynamics.

In this work, we investigate the derivative-free loss method (DFLM) \cite{DFLM} as a non-intrusive multiscale solver for fluid systems. DFLM employs a mesh-free neural network representation to approximate solutions of a class of partial differential equations (PDEs). By leveraging an equivalent stochastic formulation, the method uses stochastic walkers to estimate solution behavior without explicitly computing derivatives, and has been applied to the homogenization of heterogeneous media \cite{DFLMHomo}, where the exploration neighborhood of the walkers determines which microscale features are averaged out and which macroscale behavior is retained. DFLM captures macroscopic behavior without requiring fine-grid resolution of all relevant scales or explicit coupling between macro- and microscale models. Instead, stochastic walkers naturally encode microscale dynamics, without the need to uniformly sample the entire domain. By aggregating the averaged behavior of these walkers at collocation points, the neural network learns an effective representation of the macroscopic solution. DFLM has also been applied directly to fluid and other multiscale problems, including elliptic PDEs in perforated domains \cite{DFLMperforated} and, closest to the present work, the Navier--Stokes equations \cite{DFLMfluid}, with successful results across a range of test problems, and its theoretical properties, including the bias and convergence of the training loss with respect to the target horizon and walker size, have been analyzed in \cite{DFLManalysis}.

The application of DFLM to the Navier--Stokes equations in \cite{DFLMfluid} was demonstrated on viscous, large-scale, laminar flows, using the simplest possible way of simulating the stochastic walkers\jh{, with each walker advanced} directly from its starting point to its destination in a single step. That work explicitly identified two obstacles to extending the method toward turbulence: this single-step motion cannot capture the swirling, directional stretching that characterizes turbulent flow, and evaluating the target by randomly sampling many walkers introduces sampling error into the learned solution, which the authors noted must be corrected ``for our method to have a chance at calculating turbulence.'' The present work addresses both obstacles directly. We introduce a more refined, multi-step way of simulating the stochastic walkers (Section~\ref{sec:DFLM}), together with an analytic approximation of their local motion, developed in Section~\ref{sec:multiscale}, that captures this directional stretching exactly and replaces random sampling with a deterministic evaluation, eliminating this walker-based sampling error. We then demonstrate the resulting method on genuinely turbulent two-dimensional flows, assessing its accuracy through the macroscopic energy spectrum rather than pointwise error, consistent with our interest in capturing large-scale statistical behavior rather than resolving individual realizations exactly.

Gaussian approximations of unresolved dynamics are also widely used in multiscale and geophysical turbulence modeling \cite{MajdaTimofeyevVandenEijnden2002,GroomsMajda2013SSP,SSPparam}. In these approaches, a closure is typically calibrated from the microscale variables and then coupled explicitly to the resolved macroscale equations. DFLM takes a different, non-intrusive approach\jh{, in which} its stochastic walkers evolve directly according to the governing PDE, rather than according to a fitted surrogate. Consequently, the macroscopic Bellman target introduced in Section~\ref{sec:DFLM} is evaluated without a separate closure model or parameter-fitting step. We discuss this distinction further in Section~\ref{sec:multiscale}.

The remainder of the paper is organized as follows. Section~\ref{sec:DFLM} reviews the stochastic formulation underlying DFLM for the Navier--Stokes equations, including the Bellman target and the multi-step discretization used to simulate the stochastic walkers. Section~\ref{sec:multiscale} introduces our main technical contribution: an analytic Gaussian approximation of the walkers' local motion that resolves the anisotropy and sampling-error limitations discussed above. Section~\ref{sec:results} validates the resulting method against direct numerical simulation on two turbulent two-dimensional flows. Section~\ref{sec:discussion} discusses the implications of these results and directions for future work.

\section{Derivative-Free Loss Method for fluids}\label{sec:DFLM}
This section starts a brief review of the derivative-free loss method (DFLM) for the Navier--Stokes equations (NSE) following the ideas introduced in \cite{DFLMfluid} and \cite{DFLMHomo}. Its purpose is to formulate the problem of interest and to introduce the terminology used throughout the remainder of the paper. 

\subsection{Stochastic formulation of the Navier-Stokes equation}\label{sec:stochastic_formulation}
We consider the incompressible Navier-Stokes equation of the fluid velocity $\bm{u}(\bm{x},t)$ in a periodic domain $\Omega \subset \mathbb{R}^d$
\begin{eqnarray}
\frac{\partial \bm{u}}{\partial t} + \bm{u}\cdot \nabla\bm{u} &=& -\frac{1}{\rho} \nabla p + \nu \Delta \bm{u} + \bm{f}, ~~\text{in}~\Omega \times (0,T], \label{eq:NSE}\\
\nabla \cdot \bm{u} &=& 0 ~~\text{in}~\Omega \times (0,T].\label{eq:Incompressibility}
\end{eqnarray}
Here, $p(\bm{x},t) \in \mathbb{R}$ is the pressure, $\bm{f}$ is the force, $\rho$ is the fluid density, and $\nu$ is the viscosity with initial condition $\bm{u}(\bm{x},0)=\bm{u}_0(\bm{x})$.

DFLM is built on a martingale, or Feynman--Kac type, representation of the velocity field along a backward-in-time stochastic diffusion. Following \cite{DFLMfluid}, applying It\^o's formula to $\bm{u}(\bm{X}_s,t-s)$ along the diffusion driven by Eq.~\eqref{eq:NSE} shows that $\bm{u}(\bm{X}_s,t-s)+\int_0^s\bm{f}(\bm{X}_r,t-r)\,dr$ is a martingale, provided $\bm{X}_s$ solves the backward It\^o diffusion
\begin{equation}\label{eq:backward_sde}
d\bm{X}_{s\wedge\tau} = -\bm{u}(\bm{X}_s, t-s\wedge \tau)\,d(s\wedge \tau) + \sqrt{2\nu}\,d\bm{B}_{s\wedge \tau}, \qquad \bm{X}_0=\bm{x},
\end{equation}
where $\bm{B}_s$ is a standard Brownian motion in $\mathbb{R}^d$, $\tau=\inf \{s \mid (\bm{X}_s,t-s) \in \Omega \times \{t=0\}\}$ is a stopping time marking the first backward time at which the initial slice $t=0$ is reached, and $s\wedge \tau = \min\{s,\tau\}$; since $\Omega$ is periodic, no corresponding stopping is needed at the spatial boundary, and $\bm{X}_s$ is instead understood to evolve on the torus. Because $\tau$ is a stopping time, the Optional Stopping Theorem \cite{oksendal,ITO} preserves the martingale property of the stopped process, so that evaluating the martingale at $s=0$ and $s=\Delta t\wedge\tau$ gives the stochastic representation of the velocity field
\begin{equation}\label{eq:Stochastic_Representation}
\bm{u}(\bm{x},t) = \mathbb{E} \left[ \bm{u}(\bm{X}_{\Delta t\wedge \tau}, t-\Delta t\wedge \tau)+ \int_0^{\Delta t\wedge \tau} \bm{f}(\bm{X}_s, t-s)ds \bigg |\bm{X}_0=\bm{x} \right], ~ \forall \Delta t \in (0, t],
\end{equation}
with $\bm{X}_s$ evolving according to Eq.~\eqref{eq:backward_sde} under the incompressibility constraint Eq.~\eqref{eq:Incompressibility}. We refer to $\Delta t$ as the \emph{target horizon}\footnote{\cite{DFLMHomo} refers to $\Delta t$ as the \emph{macro-timestep}. We adopt \emph{target horizon} instead, to avoid suggesting that $\Delta t$ is itself a timestep at which the system is evolved; that role is played by the micro-timestep $\delta t$.}: it is the length of the backward window over which the stochastic representation is evaluated, not a timestep at which any system is evolved, that role being played by the micro-timestep $\delta t$ introduced in Section~\ref{sec:DFLMfluid}. For interior points with $\Delta t < \tau$, Eq.~\eqref{eq:Stochastic_Representation} reduces to the unstopped representation over the full target horizon $\Delta t$; when the backward walker reaches $t=0$ before $\Delta t$ elapses, the stopped formulation lets the initial condition enter the representation directly, allowing the walker to flow smoothly from the interior into the initial data. Section~\ref{sec:DFLMfluid} describes how this is used, together with an explicit initial-condition loss term, to enforce the initial condition in training.

Eq.~\eqref{eq:Stochastic_Representation} must be enforced together with the incompressibility constraint Eq.~\eqref{eq:Incompressibility}, i.e., the velocity field is sought in the weakly divergence-free space associated with the classical Leray--Hopf weak formulation of the NSE \cite{Temam},
\begin{equation}
V=\{\bm{v}\in H^1(\Omega;\mathbb{R}^d) : \nabla\cdot \bm{v}=0 \text{ in } \Omega\}.
\end{equation}
In the Leray--Hopf theory, the pressure is eliminated by projecting the momentum equation onto $V$, so that weak solutions are characterized entirely through divergence-free velocity fields, without reference to $p$. DFLM enforces membership in $V$ by construction, rather than by penalizing $\nabla \cdot \bm{u}$ in the loss. We introduce a neural network $\phi(\bm{x},t;\bm{\theta})$ representing a vector potential $\phi: \Omega \times [0,T] \rightarrow \mathbb{R}^{d}$ and define the velocity network as
\begin{equation}\label{eq:vector_potential}
\bm{u}(\bm{x},t;\bm{\theta})=\nabla_{\bm{x}}\times\phi(\bm{x},t;\bm{\theta}).
\end{equation}
Since $\nabla\cdot(\nabla_{\bm{x}}\times\phi)\equiv 0$ for any sufficiently smooth $\phi$, every choice of $\bm{\theta}$ automatically satisfies Eq.~\eqref{eq:Incompressibility} and lies in $V$, so that no explicit incompressibility loss term, and no separate representation of the pressure, is required. This mirrors the Leray--Hopf construction, where the pressure-free projected equation is likewise posed directly on $V$; here the same divergence-free constraint is instead realized through a differentiable parametrization suitable for gradient-based training.

\subsection{DFLM for the Navier-Stokes equation}\label{sec:DFLMfluid}
The derivative-free loss method (DFLM) trains the velocity network $\bm{u}(\bm{x},t;\bm{\theta})$ defined through Eq.~\eqref{eq:vector_potential} to satisfy the stochastic representation Eq.~\eqref{eq:Stochastic_Representation} for a fixed target horizon $\Delta t$. We refer to the conditional expectation on the right-hand side of Eq.~\eqref{eq:Stochastic_Representation}, evaluated using the current network in place of the unknown velocity field, as the \emph{Bellman target} \cite{markovreward} for the velocity field at $(\bm{x},t)$,
\begin{equation}\label{eq:bellman_target}
\mathcal{B}(\bm{x},t;\bm{\theta}) := \mathbb{E}_{\bm{X}_{0\leq s \leq \Delta t}}\left[
\bm{u}\left(\bm{X}_{\Delta t\wedge\tau},t-\Delta t\wedge\tau;\bm{\theta}\right)
+ \int_0^{\Delta t\wedge\tau} \bm{f}(\bm{X}_s,t-s)ds
\bigg |  \bm{X}_0=\bm{x}\right].
\end{equation}
The terminology follows the bootstrapped target used in temporal-difference reinforcement learning \cite{suttonbarto}. \jh{Rather} than regressing against an externally supplied label, the network is trained so that its own prediction at $(\bm{x},t)$ matches $\mathcal{B}(\bm{x},t;\bm{\theta})$, a target that is itself constructed from the network's current prediction one target horizon $\Delta t$ away. The loss function for this local martingale representation is given by
\begin{equation}\label{eq:loss}
\mathcal{L}(\bm{\theta}) = \mathbb{E}_{(\bm{x},t)}\left[
\left|\bm{u}(\bm{x},t;\bm{\theta}) - \mathcal{B}(\bm{x},t;\bm{\theta})\right|^2\right].
\end{equation}

The optimization follows a bootstrapping approach in reinforcement learning where the target value (inner expectation) and network are alternately updated. Precisely, for each $n$-th gradient descent step in optimization $\bm{\theta}_{n+1}=\bm{\theta}_n -\alpha \nabla \mathcal{L}_{n}(\bm{\theta})$ where
\begin{equation}\label{eq:loss_empirical}
\mathcal{L}_n(\bm{\theta}) = \frac{1}{N_r}\sum_{i=1}^{N_r}
\left|\bm{u}(\bm{x}_i,t_i;\bm{\theta})
- \widehat{\mathcal{B}}_i\left(\bm{\theta}_n\right)
\right|^2
\end{equation}
Here $N_r$ is the number of collocation points, $\{\bm{x}_i,t_i\}_{i=1}^{N_r}$, uniformly sampled from the spatiotemporal domain $\Omega\times (0,T]$ for an unbiased estimator of the outer expectation. The inner expectation for target evaluation is estimated in Monte-Carlo using $N_s$ \textit{stochastic walkers} \jh{$\bm{X}^{(i,j)}_s$} starting from $\bm{x}_i$, evolving according to the backward SDE Eq.~\eqref{eq:backward_sde} with $\bm{X}^{(i,j)}_0=\bm{x}_i$ for $0\leq s\leq \Delta t$ and drift given by the current network approximation $\bm{u}(\cdot,\cdot;\bm{\theta}_n)$,
\jh{
\begin{equation}\label{eq:empirical_bellman_target}
\widehat{\mathcal{B}}_i\left(\bm{\theta}_n\right)=\frac{1}{N_s}\sum \limits_{j=1}^{N_s}\bm{u}\left(\bm{X}^{(i,j)}_{\Delta t},t_i-\Delta t;\bm{\theta}_n\right) +  I_{\bm{f}}\left(\{\bm{X}_s^{(i,j)}\}_{0\leq s\leq \Delta t}\right).
\end{equation}
}
Here $N_s$ is the number of stochastic walkers simulated independently at each collocation point $(\bm{x}_i,t_i)$. The walkers \jh{$\bm{X}^{(i,j)}_s,\ldots,\bm{X}^{(i, N_s)}_s$} are independent and identically distributed (i.i.d.) realizations of the transition kernel of the backward SDE Eq.~\eqref{eq:backward_sde}, all started from the same point $\bm{x}_i$. Because $\widehat{\mathcal{B}}_i(\bm{\theta}_n)$ is a finite-sample average over these $N_s$ i.i.d. walkers, it is itself an unbiased estimator of the Bellman target $\mathcal{B}(\bm{x}_i,t_i;\bm{\theta}_n)$, with variance decreasing as $N_s$ increases. For small $N_s$, however, the resulting \emph{sampling error} can be substantial, introducing noise into the training signal that must be weighed against the added cost of simulating more walkers per collocation point. This sampling variance does not cancel out of the training loss. Because $\widehat{\mathcal{B}}_i(\bm{\theta}_n)$ enters the squared residual in Eq.~\eqref{eq:loss_empirical}, the resulting empirical loss is itself only \emph{asymptotically unbiased}, as $N_s\to\infty$, for the exact loss $\mathcal{L}(\bm{\theta})$ in Eq.~\eqref{eq:loss}. \jh{As shown in \cite{DFLManalysis}, this} bias is of order $\Delta t/N_s$ and scales with the spatial gradient of the network. Throughout this paper, \emph{sampling error} refers specifically to this walker-level noise in estimating the inner expectation at a fixed collocation point. The outer expectation retains its own Monte-Carlo error from the random choice of collocation points $\{\bm{x}_i,t_i\}$ used to estimate the loss Eq.~\eqref{eq:loss_empirical}. This outer sampling error is not addressed by the methods developed below.

\begin{figure}[t!]
    \centering
    \begin{minipage}{0.4\textwidth}
        \centering
        \includegraphics[width=\linewidth]{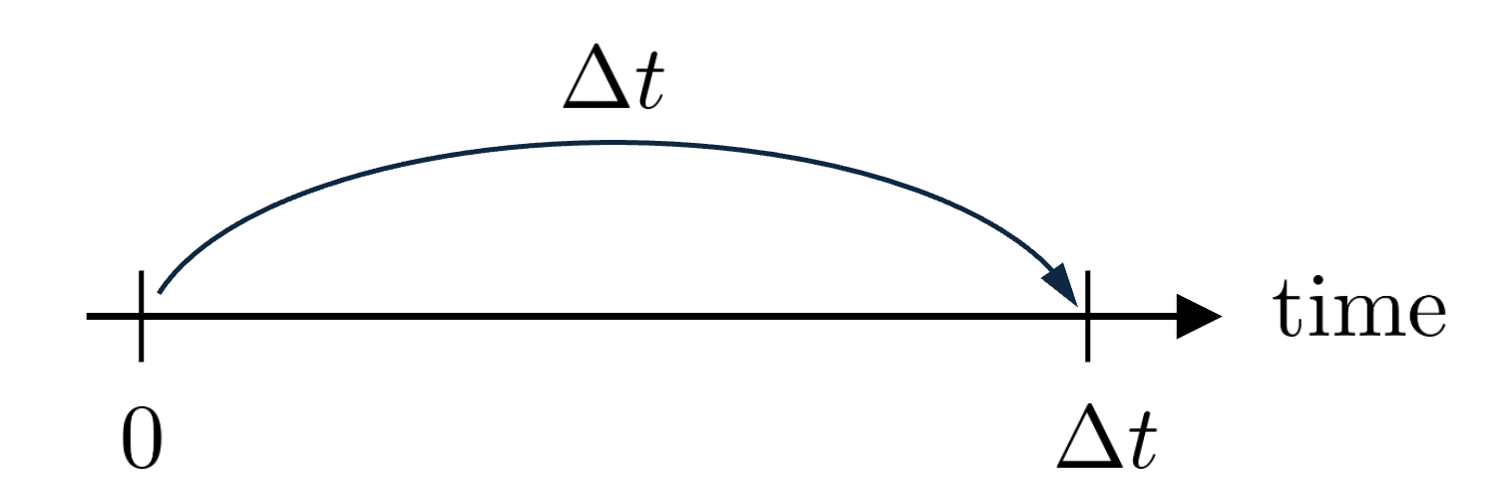}\\
        (a) one-step ($M=1$)
    \end{minipage}
    \hspace{0.06\textwidth}
    \begin{minipage}{0.4\textwidth}
        \centering
        \includegraphics[width=\linewidth]{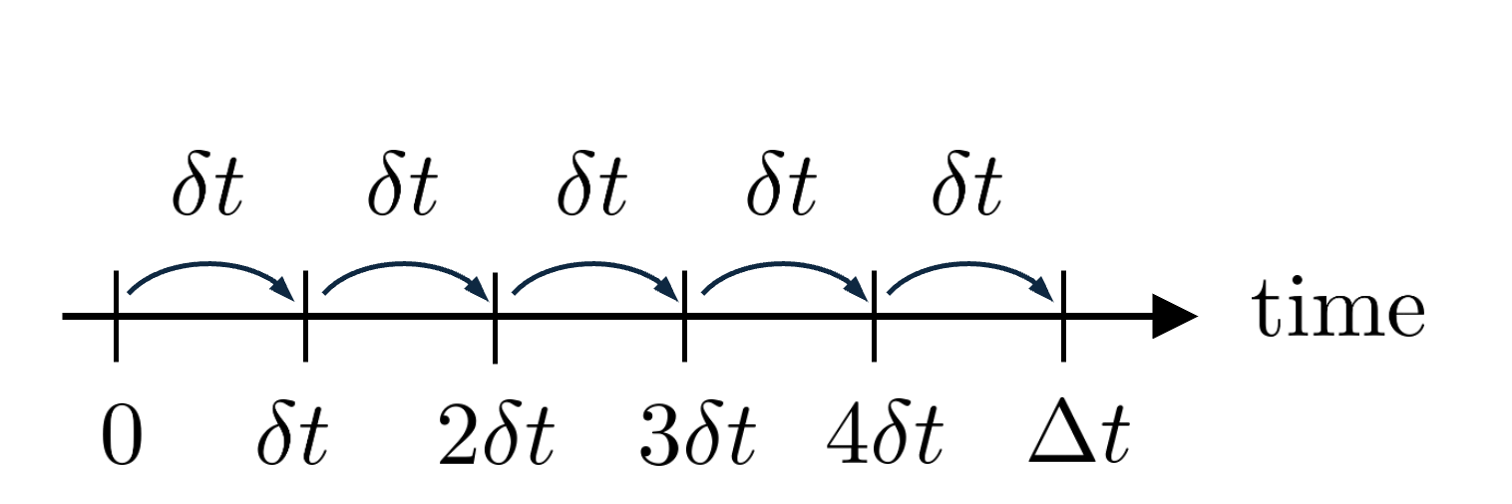}\\
        (b) multi-step ($M=5$)
    \end{minipage}
    \caption{Schematic comparison of the Euler--Maruyama discretizations of the target horizon $\Delta t$. (a) A single step directly from $t$ to $t+\Delta t$. (b) $M$ microsteps of size $\delta t$, with $\Delta t = M\delta t$.}
    \label{fig:onestep_multistep}
\end{figure}

The standard DFLM (\cite{DFLM} for elliptic problems and \cite{DFLMfluid} for fluids) evaluates the target in Eq.~\eqref{eq:empirical_bellman_target} using a single Euler--Maruyama step for computational efficiency, so that each walker is drawn directly from the one-step transition kernel of the backward SDE Eq.~\eqref{eq:backward_sde}. More generally, the target horizon $\Delta t$ can be resolved by $M$ Euler--Maruyama microsteps of size $\delta t$, such that $\Delta t = M\delta t$; \cite{DFLMHomo} introduced this multi-timestep discretization along with its analysis in resolving periodic structures, as a multiscale method, to improve the accuracy of the stochastic representation by resolving microscale features to capture macroscopic behavior. The backward SDE Eq.~\eqref{eq:backward_sde} is then discretized as
\begin{equation}\label{eq:euler_maruyama_multistep}
\bm{X}_{(m+1)\delta t} = \bm{X}_{m\delta t} - \bm{u}(\bm{X}_{m\delta t}, t-m\delta t;\bm{\theta}_n)\,\delta t + \sqrt{2\nu\delta t}\,\bm{Z}, \qquad \bm{Z}\sim\mathcal{N}(\bm{0}, \bm{I}), \quad m=0,1,\ldots,M-1,
\end{equation}
which reduces to the standard one-step update when $M=1$ and $\delta t=\Delta t$. Figure~\ref{fig:onestep_multistep} illustrates the two discretizations schematically. Here $I_{\bm{f}}$ is an approximation of the stochastic integral
\jh{
\begin{equation}
I_{\bm{f}}\left(\{\bm{X}_s^{(i,j)}\}_{0\leq s\leq \Delta t}\right) = \delta t\sum \limits_{m=0}^{M-1}\bm{f}\left(\bm{X}^{(i,j)}_{m\delta t},t_i-m\delta t\right).
\end{equation}
}
For incompressible flows, the divergence-free drift preserves volume locally, but it can still stretch fluid elements in one direction while compressing them in another. This directional stretching can make the one-step DFLM approximation produce anisotropic errors in the stochastic representation of the velocity field, as we show in the next section.

DFLM simulates both the initial and boundary conditions physically, rather than through an explicit loss term. The periodic boundary condition was already noted in Section~\ref{sec:stochastic_formulation}\jh{, where} the walker evolves on the torus, so a step that carries \jh{$\bm{X}_{s}^{(i,j)}$} outside $\Omega$ is simply mapped back in by adding the appropriate periodic vector $P_l e_l$. The initial condition is enforced in the same spirit through the stopping construction built into the backward SDE Eq.~\eqref{eq:backward_sde}\jh{. As} the walker evolves backward in time, it is stopped at $\tau$ once it reaches $t=0$ before completing the target horizon $\Delta t$, and its drift is thereafter fixed to the initial velocity $\bm{u}_0$, so that the martingale representation Eq.~\eqref{eq:Stochastic_Representation} continues to hold with $\Delta t$ replaced by the stopped horizon $\Delta t \wedge \tau$. This lets the initial data flow smoothly into the interior, allowing the trained velocity field to naturally adapt to $\bm{u}_0$. To further enforce the initial condition directly, we add a loss term using $N_a$ collocation points uniformly sampled at $t=0$,
\begin{equation}
\mathcal{L}_{\text{init}} = \frac{1}{N_a}\sum \limits_{i=1}^{N_a}|\bm{u}(\bm{x}_i, 0;\bm{\theta})- \bm{u}_0(\bm{x}_i)|^2.
\end{equation}

\section{Derivative-Free Loss Method as a multiscale method for fluids}\label{sec:multiscale}
The Bellman target $\mathcal{B}(\bm{x},t;\bm{\theta})$ in Eq.~\eqref{eq:bellman_target} gives DFLM an intrinsic averaging property that is particularly useful for multiscale problems. The conditional expectation in Eq.~\eqref{eq:bellman_target} is taken over a backward walker that explores a neighborhood of $\bm{x}$ determined by both local advection and the diffusive length scale $\sqrt{2\nu\Delta t}$ over the target horizon $\Delta t$. Matching the network to its Bellman target therefore enforces a local, physically motivated average of the velocity field over this neighborhood, rather than a pointwise fit to a single realization. A related mechanism appears in the homogenization of elliptic multiscale PDEs in \cite{DFLMHomo}\jh{, in which} the exploration neighborhood of the stochastic walkers, controlled by the time horizon, determines which microscale features are averaged out and which macroscale behavior is retained. In DFLM, the interaction between macro- and microscales is embedded in the stochastic representation itself, without introducing an explicit subgrid-scale model.

Realizing this averaging by Monte Carlo, however, is computationally costly. At each of $N_r$ collocation points, evaluating the empirical target $\widehat{\mathcal{B}}_i$ requires $N_s$ stochastic walkers, each advanced through $M$ Euler--Maruyama microsteps, giving a total cost proportional to $M N_s N_r$. Multi-step propagation can reduce the anisotropic errors of a one-step discretization, but it further increases this cost. Moreover, controlling the Monte Carlo sampling error discussed in Section~\ref{sec:DFLMfluid} requires sufficiently large $N_s$ (and typically many collocation points, $N_r$). We therefore introduce a Gaussian approximation of the stochastic process for Bellman-target evaluation, directly addressing the two obstacles \cite{DFLMfluid} identified for extending DFLM to turbulent flow. It permits an anisotropic transition covariance and removes Monte Carlo sampling variance, while avoiding the cost of large walker ensembles and, in particular, resolving within-step deformation without requiring large $M$.

The central idea of the Gaussian approximation is to locally linearize the drift over each microstep. The nonlinear stochastic dynamics are then approximated by a linear SDE, whose distribution remains Gaussian and is fully characterized by its mean and covariance. Both moments admit closed-form propagation, so they can be updated deterministically without additional sampling. The conditional expectations in the Bellman target are subsequently evaluated using Gauss--Hermite quadrature.

A key feature of this construction is its treatment of covariance. Although the Brownian forcing is isotropic, the local velocity gradient continuously deforms the diffusive distribution during each microstep. This strain-induced deformation generally makes the transition covariance anisotropic. When the initial covariance is zero, a single Euler--Maruyama step does not capture this within-step deformation, whereas the analytic covariance derived below retains it through exact propagation of the locally linearized dynamics. As demonstrated in Section~\ref{sec:results}, this distinction is important for flows with persistent shear or strain.

Gaussian closures that use microscale dynamics to inform macroscale behavior are common in multiscale and geophysical turbulence modeling \cite{MajdaTimofeyevVandenEijnden2002,GroomsMajda2013SSP, SSPparam}. Such methods generally approximate unresolved microscale variables to estimate the parameters needed to close the macroscale equations, thereby coupling the two scales through a fitted closure. DFLM takes a different approach. The Gaussian moment propagation developed below approximates the stochastic walker directly, and the resulting analytic Bellman target yields the macroscopic velocity field without a separate closure or parameter-fitting step. Furthermore, the walker SDE is derived from the governing PDE rather than inferred from data. The Gaussian approximation is therefore applied to a known stochastic process, and its deterministic Bellman target can be evaluated without Monte Carlo sampling and used directly in the training loss.

In this section, we first illustrate the limitation of the standard, one-step DFLM discretization in capturing anisotropy (Section~\ref{sec:DFLMfluid_anisotropy}). We then introduce the Gaussian approximation of the stochastic process and derive the exact mean and covariance propagation (Section~\ref{sec:Gaussian_approximation}), and describe how the resulting Gaussian is used to evaluate the Bellman target through Gauss--Hermite quadrature, comparing its computational cost against standard Monte-Carlo sampling (Section~\ref{sec:GH}). Because the mean and covariance are propagated analytically rather than through additional Euler--Maruyama microsteps, the Gaussian approximation retains anisotropy without multi-stepping while eliminating Monte-Carlo sampling error. Finally, we describe how these analytic formulas are evaluated in practice using a block matrix exponential (Section~\ref{sec:van_loan}).

\subsection{Issue of the one-step approach in capturing anisotropy}
\label{sec:DFLMfluid_anisotropy}
Many fluid systems exhibit persistent anisotropic structures rather than isotropic turbulence; for example, rotating geophysical flows organize into anisotropic zonal jets through the beta-plane mechanism \cite{Rhines1975}. Accurately resolving such anisotropy, rather than defaulting to an isotropic transition, is therefore important for the stochastic representation of the flow. For an incompressible flow, the local velocity gradient satisfies
\begin{equation}\label{eq:trace_J}
\operatorname{tr}(\nabla_{\bm{x}}\bm{u}) = \nabla\cdot\bm{u} = 0,
\end{equation}
which expresses volume preservation. Accordingly, the local flow may stretch a distribution in one direction while compressing it in another, thereby producing anisotropy.

Specializing Eq.~\eqref{eq:euler_maruyama_multistep} to $M=1$ (so $\delta t=\Delta t$) at each collocation point $\bm{x}_i$ gives the one-step update
\begin{equation}\label{eq:euler_maruyama_onestep}
\bm{X}_{\Delta t} = \bm{x}_i - \bm{u}(\bm{x}_i, t_i;\bm{\theta}_n)\,\Delta t + \sqrt{2\nu\Delta t}\,\bm{Z}, \qquad \bm{Z}\sim\mathcal{N}(\bm{0}, \bm{I}).
\end{equation}
Because every stochastic walker used to estimate the distribution of $\bm{X}_{\Delta t}$ starts from the same collocation point $\bm{x}_i$, its initial covariance is zero: $\bm{\Sigma}_0=\bm{0}$. Hence the covariance of $\bm{X}_{\Delta t}$, denoted $\bm{\Sigma}_1^{\mathrm{EM}}$, is isotropic,
\begin{equation}\label{eq:isotropic_collapse}
\bm{\Sigma}_1^{\mathrm{EM}} = 2\nu\Delta t\,\mathbf{I}.
\end{equation}
Thus, the one-step Euler--Maruyama covariance is isotropic and completely independent of the local velocity gradient. This shows that, for the one-step approach, increasing the number of Monte-Carlo samples $N_s$ can reduce the sampling error of the Euler--Maruyama estimator but cannot recover the missing within-step deformation (anisotropy), since all samples are generated from the same isotropic one-step transition kernel in Eq.~\eqref{eq:isotropic_collapse}. The discrepancy is therefore associated with the time discretization of the stochastic transition rather than with Monte-Carlo sampling error.

When multiple Euler--Maruyama microsteps are used, the isotropic covariance generated by diffusion in one microstep is subsequently deformed by the velocity gradient in later microsteps, allowing the accumulated transition covariance to become anisotropic. Refining the microstep size improves the approximation of the continuous deformation of the stochastic distribution, but also increases the computational cost. In contrast, the exact covariance propagation in Eq.~\eqref{eq:exact_covariance_general} captures the deformation induced by the locally linearized velocity field within each microstep, including when $M=1$.

\subsection{Gaussian approximation of the stochastic process}
\label{sec:Gaussian_approximation}
Consider the backward stochastic process over the $m$-th microstep, $s\in[m\delta t,(m+1)\delta t]$, and denote
\begin{equation}
\bm{\mu}_m=\mathbb{E}[\bm{X}_{m\delta t}], \qquad
\bm{\Sigma}_m=\operatorname{Cov}(\bm{X}_{m\delta t}).
\end{equation}
For a sufficiently short microstep $\delta t\ll 1$, we freeze the time dependence of the velocity at $t_m = t - m\delta t$ and linearize its spatial dependence about $\bm{\mu}_m$:
\begin{equation}\label{eq:local_velocity_linearization}
\bm{u}(\bm{x},t-s) \simeq \bm{u}_m+\mathbf{J}_m(\bm{x}-\bm{\mu}_m),
\qquad
\bm{u}_m:=\bm{u}(\bm{\mu}_m,t_m),
\qquad
\mathbf{J}_m:=\nabla_{\bm{x}}\bm{u}(\bm{\mu}_m,t_m).
\end{equation}
The stochastic process over this microstep is thus approximated by
\begin{equation}\label{eq:linearized_drift_stochastic_process}
d\bm{X}_s = \left[-\bm{u}_m-\mathbf{J}_m(\bm{X}_s-\bm{\mu}_m)\right]ds + \sqrt{2\nu}\,d\bm{B}_s,
\qquad
s\in[m\delta t,(m+1)\delta t].
\end{equation}
Eq.~\eqref{eq:linearized_drift_stochastic_process} admits the explicit solution
\begin{equation}\label{eq:linearized_sde_solution}
\bm{X}_s=\bm{\mu}_m+e^{-\mathbf{J}_m\tau}(\bm{X}_{m\delta t}-\bm{\mu}_m)
- \left(\int_0^\tau e^{-\mathbf{J}_m r}\,dr\right)\bm{u}_m
+ \sqrt{2\nu}\int_{m\delta t}^{s}e^{-\mathbf{J}_m(s-r)}\,d\bm{B}_r.
\end{equation}
where $\tau=s-m\delta t$. The solution consists of an affine transformation of $\bm{X}_{m\delta t}$ and a Gaussian stochastic integral independent of $\bm{X}_{m\delta t}$. Consequently, if $\bm{X}_{m\delta t} \sim \mathcal{N}(\bm{\mu}_m, \bm{\Sigma}_m)$, the Gaussian distribution is preserved throughout the microstep, so that $\bm{X}_{(m+1)\delta t} \sim \mathcal{N}(\bm{\mu}_{m+1},\bm{\Sigma}_{m+1})$.
Thus, the locally linearized stochastic process can be propagated deterministically by tracking only its mean $\bm{\mu}_m$ and covariance $\bm{\Sigma}_m$.

\jh{The corresponding dynamics of the mean and covariance are obtained as follows.} Taking the expectation in Eq.~\eqref{eq:linearized_drift_stochastic_process} gives
\begin{equation}\label{eq:mu_ode}
\frac{d\bm{\mu}_s}{ds} = -\bm{u}_m -\mathbf{J}_m(\bm{\mu}_s-\bm{\mu}_m),\qquad
s\in[m\delta t,(m+1)\delta t].
\end{equation}
For the covariance, define
\begin{equation}
\bm{Y}_s=\bm{X}_s-\bm{\mu}_s,
\qquad
\bm{\Sigma}_s=\mathbb{E}[\bm{Y}_s\bm{Y}_s^{\top}].
\end{equation}
Subtracting Eq.~\eqref{eq:mu_ode} from Eq.~\eqref{eq:linearized_drift_stochastic_process} gives
\begin{equation}
d\bm{Y}_s = -\mathbf{J}_m\bm{Y}_s\,ds + \sqrt{2\nu}\,d\bm{B}_s.
\end{equation}
Applying It\^o's product rule to $\bm{Y}_s\bm{Y}_s^{\top}$ and taking the expectation yields
\begin{equation}\label{eq:Sigma_ode}
\frac{d\bm{\Sigma}_s}{ds} = -\mathbf{J}_m\bm{\Sigma}_s - \bm{\Sigma}_s\mathbf{J}_m^{\top} + 2\nu\mathbf{I}, \qquad
s\in[m\delta t,(m+1)\delta t].
\end{equation}
Eqs.~\eqref{eq:mu_ode} and \eqref{eq:Sigma_ode} provide a closed system for the evolution of the first two moments of the locally linearized process. Since $\bm{u}_m$ and $\mathbf{J}_m$ are frozen over the $m$-th microstep, both equations are linear with constant coefficients \jh{and therefore admit exact closed-form solutions over the microstep.}

\jh{For} the mean, writing $\bm{y}_s=\bm{\mu}_s-\bm{\mu}_m$, Eq.~\eqref{eq:mu_ode} becomes $d\bm{y}_s/ds=-\mathbf{J}_m\bm{y}_s-\bm{u}_m$ with $\bm{y}_{m\delta t}=\bm{0}$, the same linear equation solved by the centered process in Eq.~\eqref{eq:linearized_sde_solution}; evaluating its solution at $s=(m+1)\delta t$ gives the exact mean update
\begin{equation}\label{eq:mu_evolution}
\bm{\mu}_{m+1} = \bm{\mu}_m - \left(\int_0^{\delta t} e^{-\mathbf{J}_m r}\,dr\right)\bm{u}_m.
\end{equation}
Regarding the covariance, the same integrating-factor idea applies, generalized to matrices. Define $\bm{\Phi}_s := e^{\mathbf{J}_m(s-m\delta t)}\bm{\Sigma}_s\,e^{\mathbf{J}_m^{\top}(s-m\delta t)}$; since $\mathbf{J}_m$ commutes with its own exponential, differentiating $\bm{\Phi}_s$ and substituting Eq.~\eqref{eq:Sigma_ode} for $d\bm{\Sigma}_s/ds$ gives
\begin{equation}
\frac{d\bm{\Phi}_s}{ds} = e^{\mathbf{J}_m(s-m\delta t)}\left(\mathbf{J}_m\bm{\Sigma}_s+\frac{d\bm{\Sigma}_s}{ds}+\bm{\Sigma}_s\mathbf{J}_m^{\top}\right)e^{\mathbf{J}_m^{\top}(s-m\delta t)} = 2\nu\,e^{\mathbf{J}_m(s-m\delta t)}e^{\mathbf{J}_m^{\top}(s-m\delta t)},
\end{equation}
where the $\pm\mathbf{J}_m\bm{\Sigma}_s$ terms (and their transposed counterparts) cancel. Integrating from $s=m\delta t$, where $\bm{\Phi}_{m\delta t}=\bm{\Sigma}_m$, to $s=(m+1)\delta t$ gives
\begin{equation}
e^{\mathbf{J}_m\delta t}\bm{\Sigma}_{m+1}e^{\mathbf{J}_m^{\top}\delta t} = \bm{\Sigma}_m + 2\nu\int_0^{\delta t} e^{\mathbf{J}_m\sigma}e^{\mathbf{J}_m^{\top}\sigma}\,d\sigma.
\end{equation}
Multiplying on the left by $e^{-\mathbf{J}_m\delta t}$ and on the right by $e^{-\mathbf{J}_m^{\top}\delta t}$, and substituting $r=\delta t-\sigma$ in the integral, gives the exact covariance update
\begin{equation}\label{eq:exact_covariance_general}
\bm{\Sigma}_{m+1} = e^{-\mathbf{J}_m\delta t} \bm{\Sigma}_m e^{-\mathbf{J}_m^{\top}\delta t} + \mathbf{Q}_m,
\end{equation}
where
\begin{equation}\label{eq:Qm_integral}
\mathbf{Q}_m = 2\nu\int_0^{\delta t} e^{-\mathbf{J}_m r} e^{-\mathbf{J}_m^{\top}r}\,dr.
\end{equation}
The first term in Eq.~\eqref{eq:exact_covariance_general} describes the deformation of the covariance already present at the beginning of the microstep, while $\mathbf{Q}_m$ represents the covariance generated by molecular diffusion and continuously deformed by the local velocity gradient during the microstep. 

The effect of the local velocity gradient, particularly in developing anisotropy, can be seen directly by expanding $\mathbf{Q}_m$ for small $\delta t$:
\begin{equation}\label{eq:sigma_diff_taylor}
\mathbf{Q}_m = 2\nu\delta t\,\mathbf{I} - \nu(\delta t)^2\left( \mathbf{J}_m+\mathbf{J}_m^{\top} \right) + \mathcal{O}(\delta t^3).
\end{equation}
Therefore, when $\bm{\Sigma}_m = \bm{0}$, the exact covariance of the locally linearized process satisfies
\begin{equation}\label{eq:single_step_exact_covariance}
\bm{\Sigma}_{m+1} = 2\nu\delta t\,\mathbf{I} -\nu(\delta t)^2 \left(\mathbf{J}_m+\mathbf{J}_m^{\top} \right) + \mathcal{O}(\delta t^3).
\end{equation}
This expression highlights a fundamental difference from the one-step Euler--Maruyama covariance: even when the stochastic process starts from a deterministic point ($m=0$, so $\bm{\Sigma}_0=\bm{0}$), the exact covariance already carries an anisotropic correction through the term $\left(\mathbf{J}_m+\mathbf{J}_m^{\top} \right)$.

As anticipated in Section~\ref{sec:DFLMfluid_anisotropy}, this stretching-and-compressing mechanism can be made precise by decomposing the velocity gradient into its symmetric and antisymmetric parts,
\begin{equation}\label{eq:J_decomposition}
\mathbf{J}_m = \mathbf{S}_m+\mathbf{\Omega}_m,
\qquad
\mathbf{S}_m =\frac{1}{2} \left( \mathbf{J}_m+\mathbf{J}_m^{\top} \right),
\qquad
\mathbf{\Omega}_m = \frac{1}{2} \left( \mathbf{J}_m-\mathbf{J}_m^{\top} \right),
\end{equation}
where $\mathbf{S}_m$ is the local rate-of-strain tensor and $\mathbf{\Omega}_m$ the local rotational component. \jh{ Eq.~\eqref{eq:single_step_exact_covariance} can then be written as}
\begin{equation}\label{eq:single_step_strain}
\bm{\Sigma}_{m+1} = 2\nu\delta t \left( \mathbf{I}-\delta t\,\mathbf{S}_m \right) + \mathcal{O}(\delta t^3).
\end{equation}
Thus, the leading-order departure from the isotropic Brownian covariance is governed by the local rate of strain, not the rotation. For the two-dimensional incompressible flow, $\operatorname{tr}(\mathbf{S}_m) = \operatorname{tr}(\mathbf{J}_m) = 0$, so $\mathbf{S}_m$ has eigenvalues $\lambda_m$ and $-\lambda_m$, giving leading-order covariance eigenvalues
\begin{equation}\label{eq:covariance_eigenvalues}
2\nu\delta t(1-\delta t\lambda_m)
 \qquad
 \text{and}
 \qquad
2\nu\delta t(1+\delta t\lambda_m),
\end{equation}
which are generally unequal. \jh{This corresponds to stretching of the diffusive distribution along one principal strain direction while compression along the other.} The rotational component $\mathbf{\Omega}_m$ does not appear in Eq.\eqref{eq:single_step_strain}, consistent with the fact that rigid rotation alone cannot deform an isotropic distribution into an anisotropic one. In contrast, the one-step Euler--Maruyama covariance in Eq.~\eqref{eq:isotropic_collapse} contains no dependence on $\mathbf{S}_m$ and remains isotropic regardless of the local strain.

\subsection{Gauss-Hermite quadrature for Bellman target}\label{sec:GH}
Using the locally linearized dynamics described above, the terminal distribution of the stochastic process is approximated by
\begin{equation}
\bm{X}_{\Delta t} \sim \mathcal{N}(\bm{\mu}_M,\bm{\Sigma}_M).
\end{equation}
We use this Gaussian approximation to replace the Monte-Carlo evaluation of the terminal contribution to the Bellman target with a deterministic Gauss--Hermite quadrature \cite{GolubWelsch1969}.

Let $\mathbf{L}_M$ be the Cholesky decomposition of $\bm{\Sigma}_M = \mathbf{L}_M\mathbf{L}_M^{\top}$. Then the terminal expectation is approximated as
\begin{equation}\label{eq:GH_terminal}
\mathbb{E}\left[ \bm{u}(\bm{X}_{\Delta t},t-\Delta t) \bigg| \bm{X}_0=\bm{x} \right]
\simeq
\sum_{j=1}^{N_{\mathrm{GH}}} w_j \bm{u} \left( \bm{\mu}_M+\mathbf{L}_M\bm{\xi}_j, t-\Delta t \right),
\end{equation}
where $\{\bm{\xi}_j, w_j\}_{j=1}^{N_{\mathrm{GH}}}$ denote the Gauss--Hermite quadrature nodes and weights associated with the standard Gaussian distribution.

For the forcing contribution, we use a left-point approximation along the propagated mean trajectory:
\begin{equation}\label{eq:force_mean}
\mathbb{E}\left[ \int_0^{\Delta t} \bm{f}(\bm{X}_s,t-s)ds \bigg| \bm{X}_0=\bm{x} \right]
\simeq
\delta t \sum_{m=0}^{M-1} \bm{f}(\bm{\mu}_m,t-m\delta t).
\end{equation}
Therefore, the deterministic approximation of the Bellman target is given by
\begin{equation}\label{eq:GH_target}
\sum_{j=1}^{N_{\mathrm{GH}}} w_j \bm{u} \left( \bm{\mu}_M+\mathbf{L}_M\bm{\xi}_j, t-\Delta t \right) + 
\delta t \sum_{m=0}^{M-1} \bm{f}(\bm{\mu}_m,t-m\delta t).
\end{equation}
The standard Monte-Carlo construction requires $N_s$ stochastic walkers with $M$ microsteps for each collocation point, resulting in a cost proportional to $N_s M$ for walker propagation. In contrast, the proposed method propagates a single pair of moments $(\bm{\mu}_m, \bm{\Sigma}_m)$ through the $M$ microsteps and \jh{applies Gauss–Hermite quadrature only once, at the terminal distribution characterized by $(\bm{\mu}_M,\bm{\Sigma}_M)$. Therefore, the quadrature requires a fixed number $N_{\mathrm{GH}}=q^d$ of function evaluations, independent of $M$.} In the two-dimensional problems considered in this work, $N_{\mathrm{GH}} = q^2$. The resulting target is deterministic and eliminates Monte-Carlo sampling variance while retaining, through the exact covariance propagation, the local anisotropic deformation generated by the velocity gradient.

\subsection{Computation of the analytic formula of the mean and covariance}
\label{sec:van_loan}
We emphasize that Eq.~\eqref{eq:single_step_strain} is used only to identify the leading-order mechanism responsible for this deformation. In the current work, the covariance and the mean are not approximated by a truncated expansion. Instead, Eqs.~\eqref{eq:mu_evolution} and \eqref{eq:exact_covariance_general} are evaluated directly, retaining the full finite-$\delta t$ dependence on the locally linearized velocity gradient $\mathbf{J}_m$.

Both matrix integrals can be evaluated following Van Loan's method \cite{VanLoan1978} without numerical integration, by augmenting the generator so that the desired integral appears as a block of a single matrix exponential. For the mean, we augment the generator with the constant drift $\bm{u}_m$, defining the $(d+1)\times(d+1)$ matrix $\mathbf{A}_m$ and its exponential
\begin{equation}
\mathbf{A}_m = \begin{bmatrix}-\mathbf{J}_m & \bm{u}_m \\ \bm{0}^\top & 0\end{bmatrix},
\qquad
\Phi(t):=\exp(\mathbf{A}_mt)=\begin{bmatrix}F(t) & g(t) \\ \bm{0}^\top & h(t)\end{bmatrix}.
\end{equation}
Differentiating $\Phi$ against its generator and matching blocks gives
\begin{equation}
\dot{F}=-\mathbf{J}_mF,\qquad \dot{h}=0, \qquad\text{and}\qquad \dot{g}=-\mathbf{J}_mg+\bm{u}_m,
\end{equation}
with $F(0)=\mathbf{I}$, $h(0)=1$, $g(0)=\bm{0}$. Solving these by an integrating factor gives $F(t)=e^{-\mathbf{J}_mt}$, $h(t)\equiv1$, and $g(t)=\left(\int_0^t e^{-\mathbf{J}_mr}\,dr\right)\bm{u}_m$, which is exactly the integral in Eq.~\eqref{eq:mu_evolution} evaluated at $t=\delta t$. \jh{Hence, the mean is updated as $\bm{\mu}_{m+1}=\bm{\mu}_m-\bm{g}_m$, where $\bm{g}_m$ is obtained from the upper-right block of $\Phi(t)$ evaluated at $t=\delta t$,
\begin{equation}\label{eq:block_mean}
\Phi(\delta t) = 
\begin{bmatrix} F(\delta t) & g(\delta t) \\ \bm{0}^\top & 1\end{bmatrix}
=
\begin{bmatrix}\mathbf{F}_m & \bm{g}_m \\ \bm{0}^\top & 1\end{bmatrix}.
\end{equation}
}

The same idea applies to the covariance integral $\mathbf{Q}_m$ in Eq.~\eqref{eq:Qm_integral}, \jh{which can be evaluated through a second block matrix exponential.} At each microstep, after evaluating $\mathbf{J}_m$, we construct the augmented $2d\times 2d$ matrix $\mathbf{B}_m$ and its exponential
\begin{equation}
\mathbf{B}_m=\begin{bmatrix}
-\mathbf{J}_m & 2\nu\mathbf{I}\\
\mathbf{0} & \mathbf{J}_m^{\top}
\end{bmatrix},
\qquad
\Psi(t):=\exp(\mathbf{B}_mt).
\end{equation}
By the same argument as above,
\begin{equation}
\Psi(\delta t)=\begin{bmatrix}
\mathbf{F}_m & \mathbf{C}_m\\
\mathbf{0} & \mathbf{F}_m^{-\top}
\end{bmatrix},
\end{equation}
recovering the same propagator $\mathbf{F}_m=e^{-\mathbf{J}_m\delta t}$ already used for the mean in Eq.~\eqref{eq:block_mean}. The upper-right block $\mathbf{C}_m$, together with $\mathbf{F}_m$, determines the diffusion covariance as
\begin{equation}\label{eq:Qm_block}
\mathbf{Q}_m = \mathbf{C}_m\mathbf{F}_m^{\top}.
\end{equation}
Substituting Eq.~\eqref{eq:Qm_block} into Eq.~\eqref{eq:exact_covariance_general} gives the covariance update
\begin{equation}\label{eq:sigma_exact_update}
\bm{\Sigma}_{m+1} = \mathbf{F}_m\bm{\Sigma}_m\mathbf{F}_m^{\top} + \mathbf{C}_m\mathbf{F}_m^{\top}.
\end{equation}
Thus, evaluating both the mean and covariance updates at each microstep requires only the local velocity gradient $\mathbf{J}_m$ and drift $\bm{u}_m$, construction of the two augmented generators, and computation of their matrix exponentials; no matrix inversion or separate numerical quadrature of Eq.~\eqref{eq:mu_evolution} or Eq.~\eqref{eq:Qm_integral} is required.

\section{Numerical Results}\label{sec:results}
We validate DFLM as a non-intrusive multiscale solver for the incompressible turbulent Navier--Stokes equation with a wide range of active scales, focusing on the two mechanisms introduced in Section~\ref{sec:multiscale} for resolving anisotropic transitions: multiple Euler--Maruyama microsteps (Section~\ref{sec:DFLMfluid_anisotropy}) and the analytic Gaussian approximation (Section~\ref{sec:Gaussian_approximation}). We consider two random-forcing regimes, broadband forcing (Example~1, Section~\ref{sec:broadband}) and narrowband forcing (Example~2, Section~\ref{sec:narrowband}), and for each compare three constructions of the velocity field: a reference direct numerical simulation (DNS), the standard Monte-Carlo DFLM target of Section~\ref{sec:DFLMfluid}, and the Gaussian-approximation target of Section~\ref{sec:GH}; figures below label these DFLM MC and DFLM GH, respectively.

The reference DNS solution is computed with a $512\times512$ pseudo-spectral method with two-thirds dealiasing, advanced in time with a fourth-order exponential time-differencing Runge--Kutta scheme (ETDRK4) at time step $\delta t_{\mathrm{DNS}}=5\times10^{-4}$.

The velocity network $\bm{u}(\bm{x},t;\bm{\theta})$ and the vector-potential network of Eq.~\eqref{eq:vector_potential} use a multilayer perceptron (MLP) with seven hidden layers of $200$ neurons each, employing the swish activation in both examples. The empirical loss of Eq.~\eqref{eq:loss_empirical} is evaluated at $N_r=5000$ collocation points per iteration in the spatiotemporal domain. To impose the initial condition explicitly in addition to the implicit implementation, we sample $N_a=2000$ collocation points at $t=0$. The standard Monte-Carlo target of Section~\ref{sec:DFLMfluid} uses $N_s=300$ stochastic walkers, and the Gaussian-approximation target of Section~\ref{sec:GH} uses $N_{\mathrm{GH}}=9 \times 9$ Gauss--Hermite nodes. Both networks are trained using the Adam optimizer with $\beta_1=0.99$ and $\beta_2=0.99$, with the learning rate decayed exponentially from an initial value $\alpha=5\times10^{-4}$ by a factor $\gamma=0.9$ every $5000$ iterations, for a total of $10^5$ training iterations. At each iteration, the target is evaluated and held fixed, and the network parameters are updated from $\bm{\theta}_{n-1}$ to $\bm{\theta}_n$ through $K=3$ gradient steps against this fixed target, ensuring the network adequately fits each bootstrapped target before it is refreshed:
\begin{equation}
\bm{\theta}_{n-1}^{(k+1)}=\bm{\theta}_{n-1}^{(k)}-\alpha \nabla \tilde{\mathcal{L}}_n\left(\bm{\theta}_{n-1}^{(k)}\right),~k=0,1,\cdots,K-1, {\quad} \bm{\theta}_{n-1}^{(0)}=\bm{\theta}_{n-1},~\bm{\theta}_{n-1}^{(K)} = \bm{\theta}_n.
\end{equation}
For the target horizon, $\Delta t$, Example~1 uses $\Delta t=10^{-4}$, and Example~2 uses $\Delta t=5\times10^{-4}$. In the multistep comparison, both examples use $M=5$ microsteps over the same target horizon as their respective one-step cases, corresponding to $\delta t=2\times10^{-5}$ in Example~1 and $\delta t=10^{-4}$ in Example~2, respectively.

\begin{figure}[t!]
    \centering
    \begin{minipage}{0.48\textwidth}
        \centering
        \includegraphics[width=\linewidth]{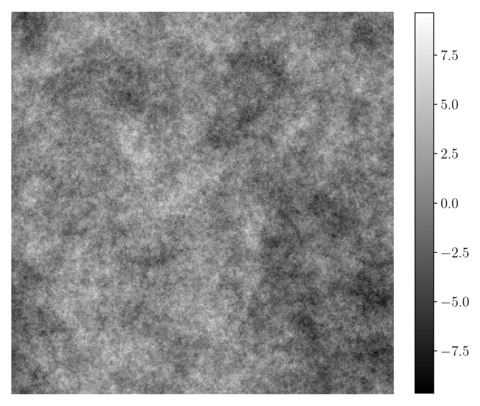}\\
        (a) broadband
    \end{minipage}
    \hfill
    \begin{minipage}{0.48\textwidth}
        \centering
        \includegraphics[width=\linewidth]{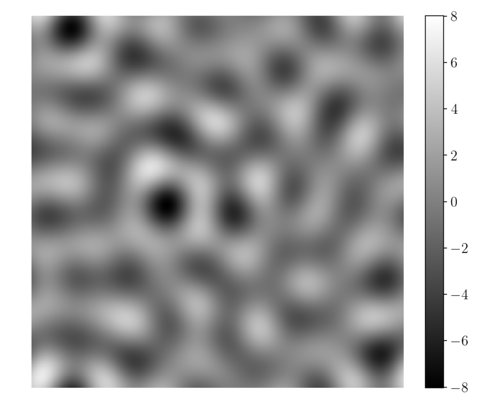}\\
        (b) narrowband
    \end{minipage}
    \caption{Vorticity forcing for the two random-forcing types.}
    \label{fig:rand_vorticity_force}
\end{figure}

In both examples below, the domain is $\Omega=[0,2\pi]^2$, the viscosity is $\nu=0.005$, and the time interval is $[0,T]$ with $T=5$. The flow is driven by the common initial condition and a time-independent, divergence-free random body force $\bm{f}=(f_x,f_y)$. Figure~\ref{fig:rand_vorticity_force} shows a representative realization of the vorticity source for each forcing type; further details are given in the corresponding experiment descriptions.

Both examples share the same initial vorticity field $\omega_0$, constructed independently of the forcing. In Fourier space, $\omega_0$ has a flat (``square'') amplitude on an annulus of intermediate wavenumbers and vanishes elsewhere,
\begin{equation}\label{eq:ic_construction}
\hat{\omega}_0(\mathbf{k}) =
\begin{cases}
e^{i\theta_{\mathbf{k}}}, & k_{\min} \leq |\mathbf{k}| \leq k_{\max},\\[2pt]
0, & \text{otherwise},
\end{cases}
\qquad
\theta_{\mathbf{k}} \sim \mathrm{Unif}(0,2\pi),
\end{equation}
with an independently generated random phase $\theta_{\mathbf{k}}$ on each active mode. Hermitian symmetry is imposed so that $\omega_0$ is real-valued, and the resulting field is rescaled to a prescribed root-mean-square (RMS) amplitude, following the same normalization used for the forcing below. We choose $k_{\min}=4$, $k_{\max}=8$, and an RMS amplitude of $3.0$, confining the initial vorticity to a band of intermediate wavenumbers.

The two examples differ only in the spectral bandwidth of this forcing. Broadband forcing (Section~\ref{sec:broadband}) spreads energy with independent phases across roughly two decades of wavenumbers, giving the injected field a short, scale-mixed spatial coherence length. Narrowband forcing (Section~\ref{sec:narrowband}) instead concentrates energy in a thin annulus of relative width $\Delta k_f/k_f\approx1/6$, giving it a well-defined spatial coherence length that is further organized, through the two-dimensional inverse energy cascade, into large-scale coherent vortices with persistent strain between them. As a result, the broadband example produces a nearly isotropic stochastic transition, while the narrowband example develops substantial, persistent anisotropy\jh{, for which resolving the exact transition covariance becomes important.}

We assess each method through three complementary diagnostics: instantaneous streamline snapshots, instantaneous energy-spectrum snapshots, and the energy spectrum averaged over the final part of the simulation, $t\in[4.5,5]$, by which point the flow has reached a statistically stationary state. Our focus throughout is on how accurately each method captures the macroscopic, large-to-intermediate-scale energy spectrum, rather than the fine-scale dissipation range.

A distinctive feature of DFLM is how these snapshots are obtained. Because the network $\bm{u}(\bm{x},t;\bm{\theta})$ is trained on collocation points sampled from the entire spatiotemporal domain $\Omega\times(0,T]$, rather than advanced step by step in time, the solution at every reported time is available as soon as training is complete\jh{. Every} snapshot shown below is an evaluation of a single trained network at a different value of $t$, not the output of a sequential time-marching scheme.

\subsection{Example 1: Broadband random forcing}\label{sec:broadband}

To guarantee incompressibility automatically, we generate $\bm{f}$ from a scalar vorticity source $f_\omega$ through a stream function $\psi_f$, setting $f_\omega=\Delta\psi_f$ and $\bm{f} = \left( \partial\psi_f/\partial x_2, -\partial\psi_f/\partial x_1 \right)$; in Fourier space this gives
\begin{equation}
\hat{f}_x(\mathbf{k}) = -\frac{i k_1}{|\mathbf{k}|^2}\hat{f}_\omega(\mathbf{k}),
\qquad
\hat{f}_y(\mathbf{k}) = \frac{i k_2}{|\mathbf{k}|^2}\hat{f}_\omega(\mathbf{k}).
\end{equation}
For the broadband case, $f_\omega$ is a Gaussian random field with Fourier amplitudes that decay algebraically with the wavenumber,
\begin{equation}\label{eq:broadband_forcing}
\hat{f}_{\omega}(\mathbf{k}) = \frac{ Z_{\mathbf{k}}^{r} + iZ_{\mathbf{k}}^{i} }{ \sqrt{2}} |\mathbf{k}|^{-s},
\qquad
Z_{\mathbf{k}}^{r}, Z_{\mathbf{k}}^{i} \sim \mathcal{N}(0,1),
\end{equation}
with Hermitian symmetry imposed so that $f_\omega$ is real-valued, and the resulting field rescaled to a prescribed RMS amplitude. We choose $s=1$, giving a broadband forcing whose energy is spread over a wide range of spatial wavenumbers. Figure~\ref{fig:rand_vorticity_force}(a) shows a representative realization of this vorticity forcing.

\begin{figure}[t!]
\centering
\includegraphics[width=1\textwidth]{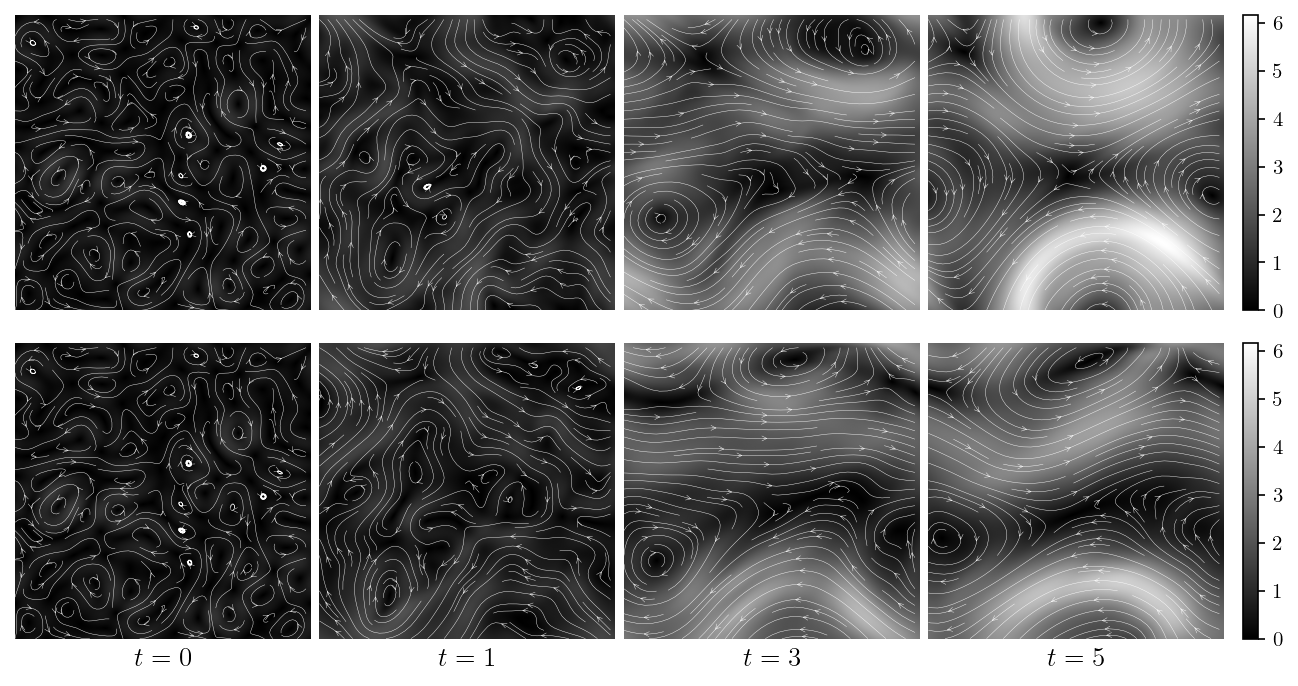}
\caption{Broadband-forcing streamline snapshots. Top: DNS. Bottom: One-step DFLM GH.}\label{fig:ex1_streamline_snapshots}
\end{figure}

\begin{figure}[t!]
\centering
\includegraphics[width=1\textwidth]{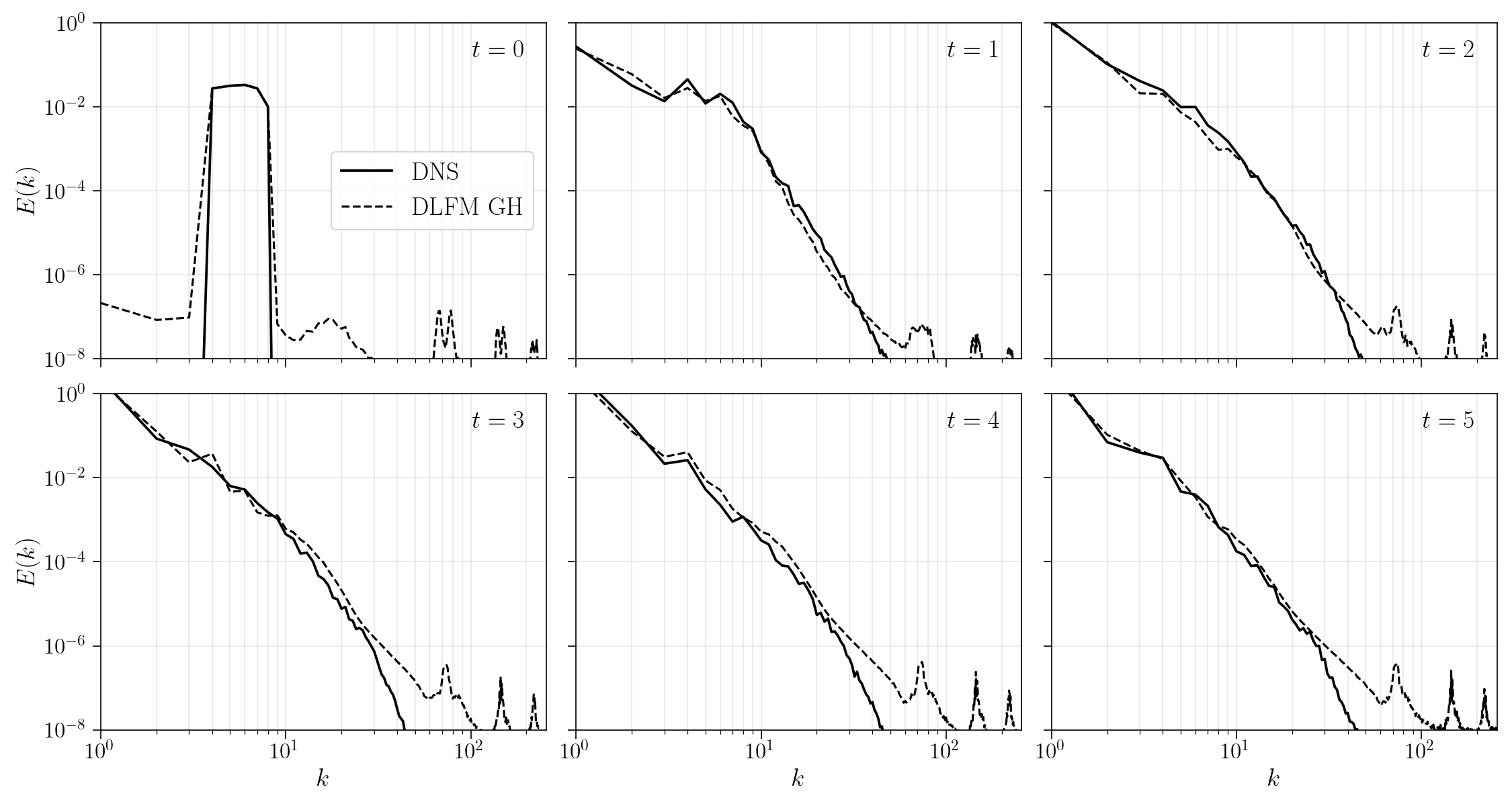}
\caption{Broadband-forcing spectrum snapshots of DNS and one-step DFLM GH.}\label{fig:ex1_seq_spectrum}
\end{figure}

 Because energy is injected simultaneously across many scales with independent phases, the resulting DNS spectrum decays monotonically from the largest scales and remains close to isotropic throughout the simulation, as seen in the streamline snapshots of Figure~\ref{fig:ex1_streamline_snapshots} and the spectrum sequence of Figure~\ref{fig:ex1_seq_spectrum}. \jh{By $t=1$, the influence of the initial condition on the energy spectrum has effectively vanished,  well before the diagnostic window $t\in[4.5,5]$ considered below.}

\begin{figure}[t!]
    \centering
    \begin{minipage}{0.48\textwidth}
        \centering
        \includegraphics[width=\linewidth]{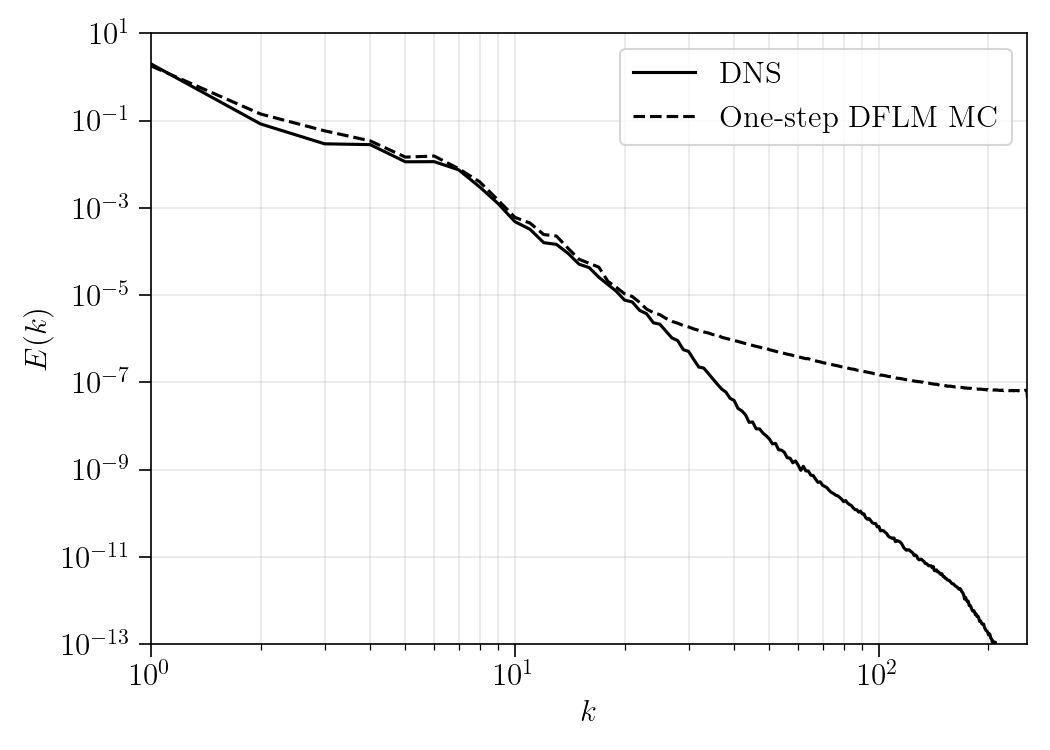}
        (a) One-step DFLM MC
    \end{minipage}
    \hfill
    \begin{minipage}{0.48\textwidth}
        \centering
        \includegraphics[width=\linewidth]{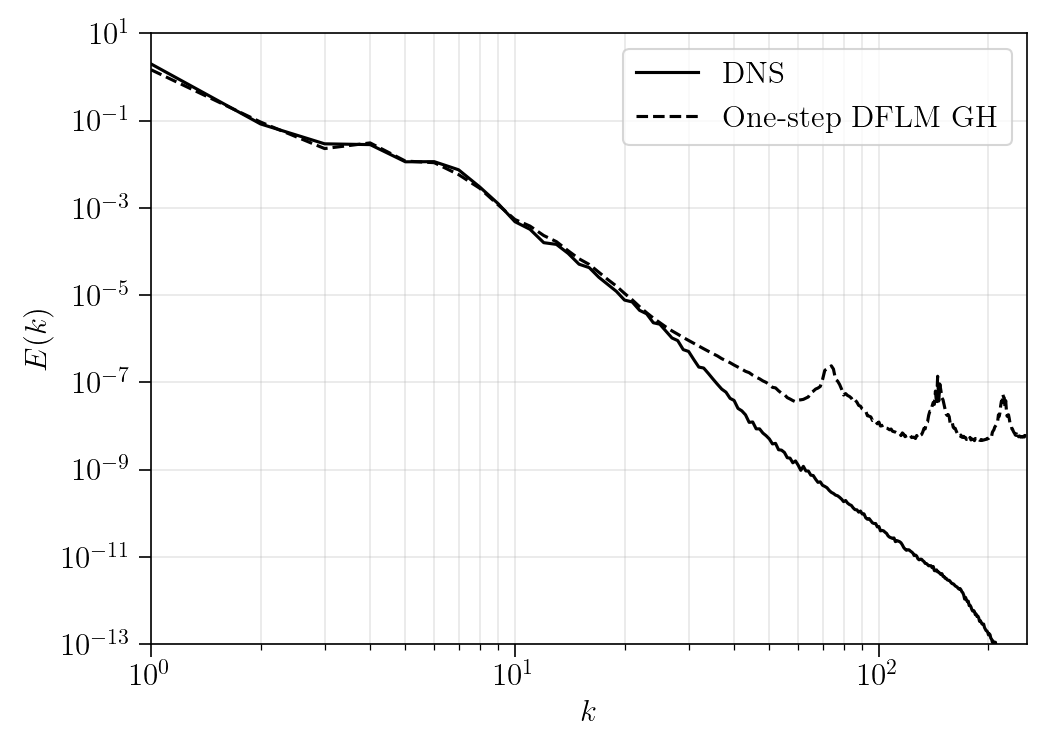}
        (b) One-step DFLM GH
    \end{minipage}\\
    \begin{minipage}{0.48\textwidth}
        \centering
        \includegraphics[width=\linewidth]{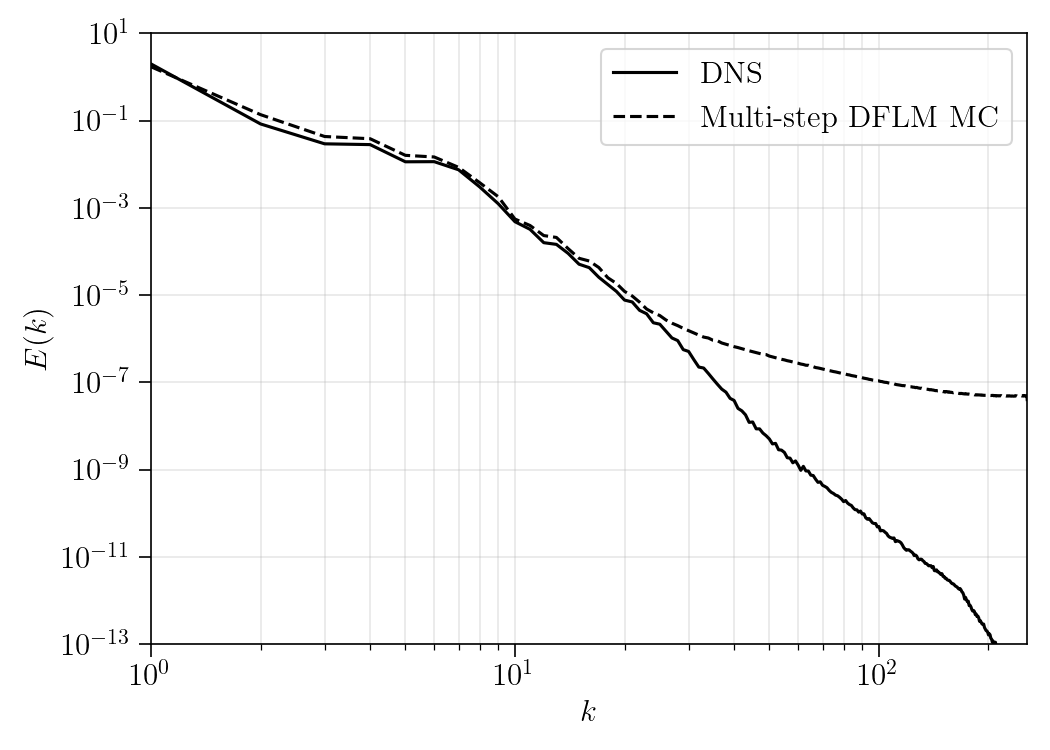}
        (c) Multistep DFLM MC
    \end{minipage}
    \hfill
    \begin{minipage}{0.48\textwidth}
        \centering
        \includegraphics[width=\linewidth]{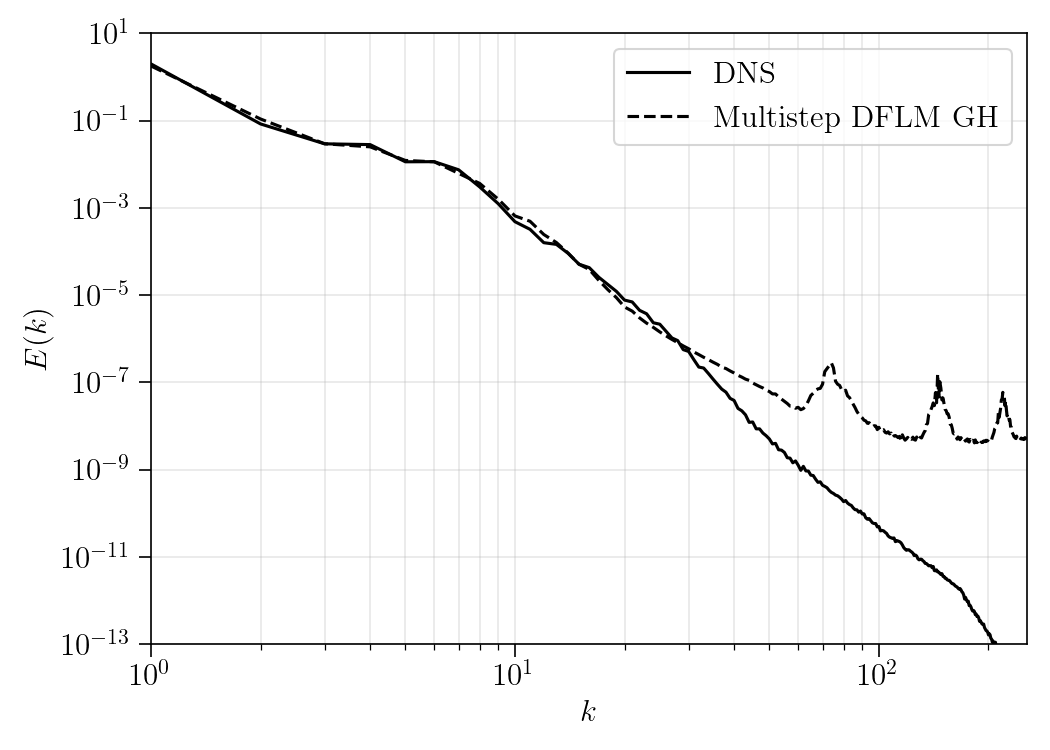}
        (d) Multistep DFLM GH
    \end{minipage}
    \caption{Broadband-forcing time-averaged energy spectra over $t\in[4.5,5]$. The multistep results uses $M=5$ for both DFLM MC and DFLM GH.}
    \label{fig:ex1_avg_spectrum}
\end{figure}

Figure~\ref{fig:ex1_avg_spectrum} compares the time-averaged spectra obtained with the Monte-Carlo target (left column) and the Gaussian-approximation target (right column), using both one-step and multistep approaches, against DNS. Both methods reproduce the DNS spectrum closely up to $k\approx20$--$25$. At higher wavenumbers, the spectra develop a flattened tail rather than following the continued algebraic decay of DNS, reflecting the finite sampling and quadrature budgets rather than a systematic bias. Because the flow is nearly isotropic in this example, the two target constructions agree closely with each other and with DNS. Using multiple microsteps ($M=5$) produces results essentially indistinguishable from the one-step targets, since the target horizon $\Delta t$ is sufficiently short that within-step deformation is negligible. This agreement over the macroscopic, large-to-intermediate-scale range provides a central validation of DFLM as a multiscale solver\jh{. Although} neither target resolves the flow on a fine computational mesh, both recover the energy-containing scales of the fully resolved DNS through the intrinsic averaging property of the Bellman target described in Section~\ref{sec:multiscale}.

The comparable results of the Monte-Carlo and Gaussian-approximation targets are consistent with the bandwidth argument above: because broadband forcing has no single dominant scale, the local strain field is spatially incoherent and remains small relative to the diffusive covariance over the target horizon, so the isotropic one-step transition already provides an accurate approximation of the true, only weakly anisotropic transition kernel. In this near-isotropic regime, the Gaussian approximation therefore offers little accuracy advantage over Monte-Carlo, but it remains preferable on efficiency grounds without the sampling variance that Monte-Carlo requires additional walkers to control. This efficiency advantage becomes essential, rather than incidental, once anisotropy is significant, as in Example~2 below.

\subsection{Example 2: Narrowband random forcing}\label{sec:narrowband}
The narrowband example uses the same domain, viscosity, and time interval as above, but changes the spectral structure of the forcing. The divergence-free force $\bm{f}$ is again constructed from a vorticity source $f_\omega$ through the stream-function relation of Section~\ref{sec:broadband}, but instead of the broadband spectrum in Eq.~\eqref{eq:broadband_forcing}, $f_\omega$ is now confined to a thin annulus in Fourier space around a characteristic wavenumber $k_f$,
\begin{equation}\label{eq:narrowband_forcing}
\hat{f}_{\omega}(\mathbf{k}) =
\begin{cases}
A\,e^{i\phi_{\mathbf{k}}}, & \left|\,|\mathbf{k}|-k_f\,\right| \leq \Delta k_f,\\[2pt]
0, & \text{otherwise},
\end{cases}
\qquad
\phi_{\mathbf{k}} \sim \mathrm{Unif}(0,2\pi),
\end{equation}
with Hermitian symmetry again imposed so that $f_\omega$ is real-valued, and the field rescaled to the same prescribed RMS amplitude. We choose $k_f=6$, $\Delta k_f=1$, and $A=2.0$: rather than spreading energy over a broad range of wavenumbers as in Example 1, this forcing injects energy at a single scale. Figure~\ref{fig:rand_vorticity_force}(b) shows a representative realization of this narrowband vorticity forcing.

\begin{figure}[t!]
\centering
\includegraphics[width=1\textwidth]{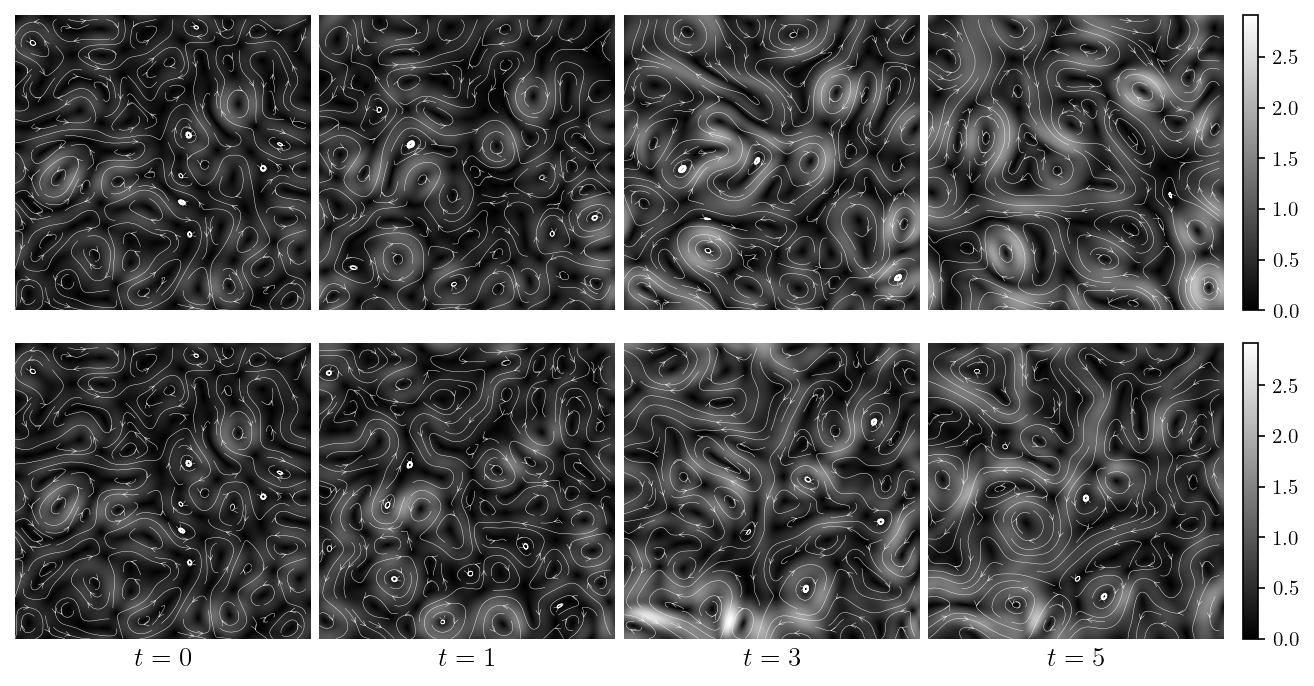}
\caption{Narrowband-forcing streamline snapshots. Top: DNS. Bottom: One-step DFLM GH.}\label{fig:ex2_streamline_snapshots}
\end{figure}

\begin{figure}[t!]
\centering
\includegraphics[width=1\textwidth]{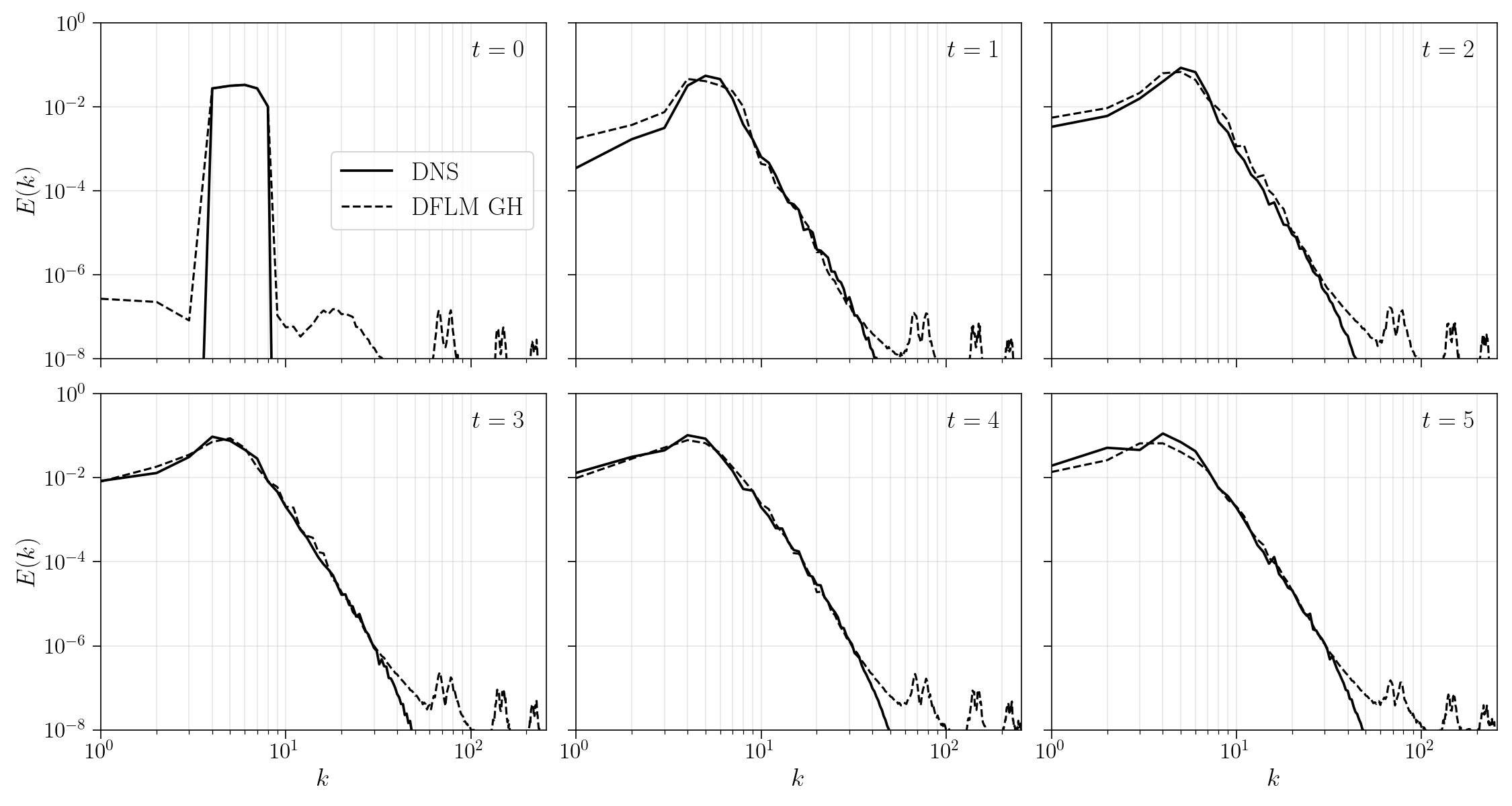}
\caption{Narrowband-forcing spectrum snapshots of DNS and one-step DFLM GH.}\label{fig:ex2_seq_spectrum}
\end{figure}

 Unlike the broadband case, the resulting DNS spectrum is not monotonic\jh{. It} rises to a peak near the injection scale $k_f$, the integral (largest) scale of the flow, then falls through an inertial range before reaching the dissipative range at high $k$, reflecting the inverse energy cascade discussed above. Together with the streamline snapshots of Figure~\ref{fig:ex2_streamline_snapshots} and the spectrum sequence of Figure~\ref{fig:ex2_seq_spectrum}, this turbulent spectral shape shows visibly more organized, anisotropic flow structure than Example 1.

\begin{figure}[t!]
    \centering
    \begin{minipage}{0.48\textwidth}
        \centering
        \includegraphics[width=\linewidth]{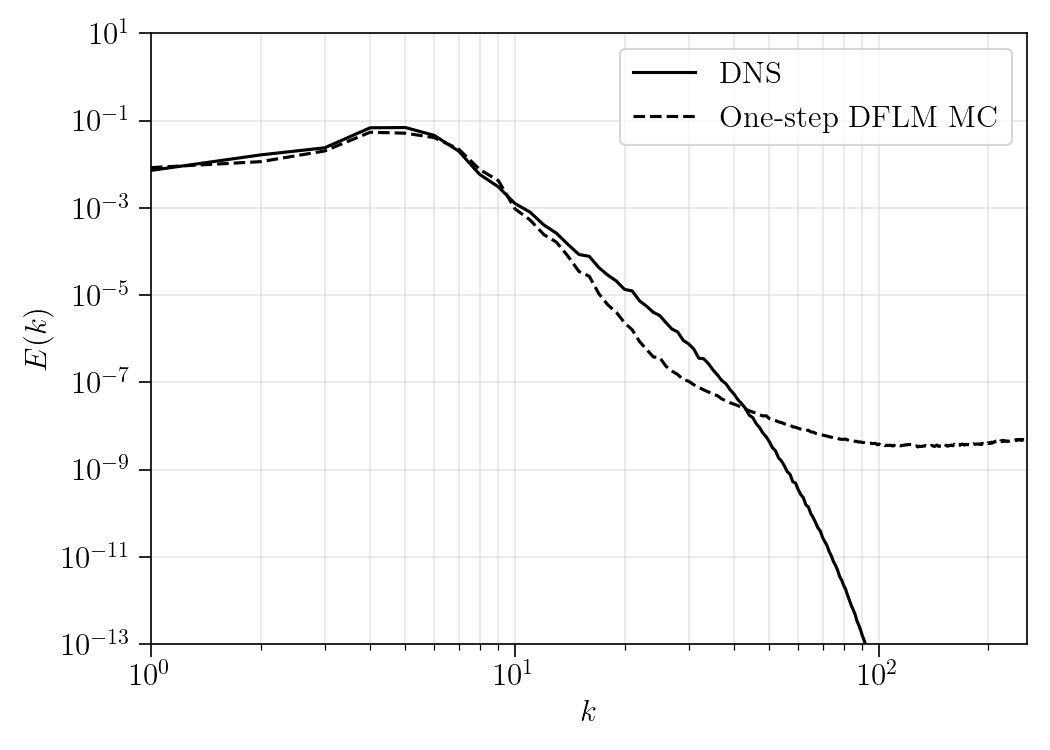}
        (a) One-step DFLM MC
    \end{minipage}
    \hfill
    \begin{minipage}{0.48\textwidth}
        \centering
        \includegraphics[width=\linewidth]{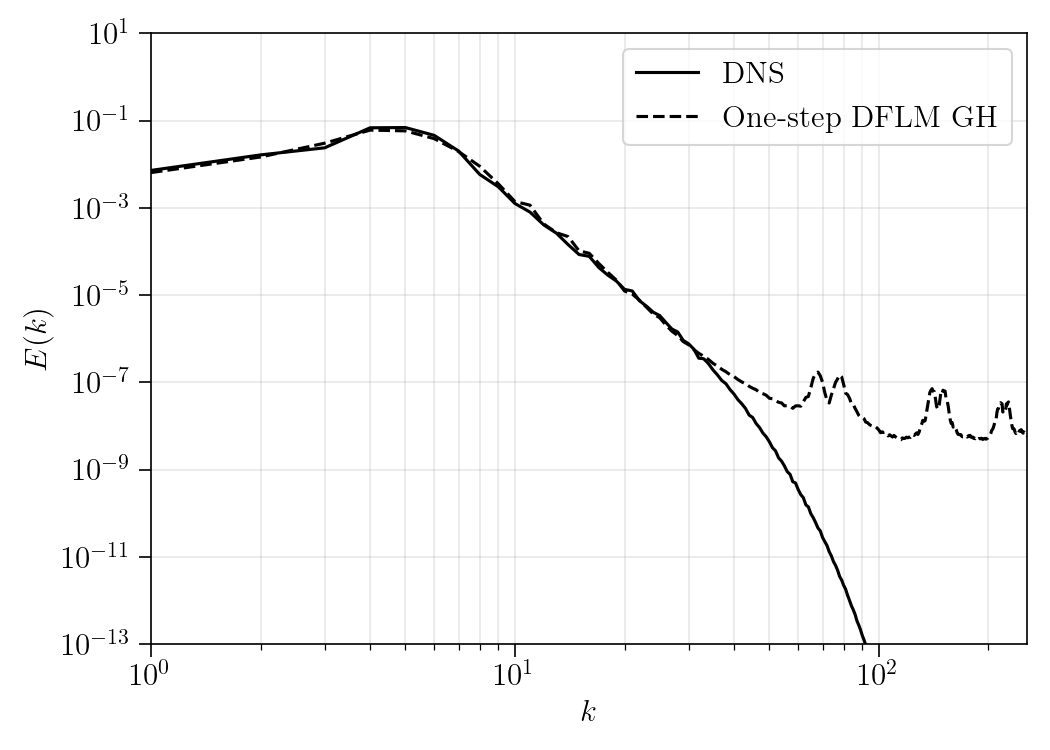}
        (b) One-step DFLM GH
    \end{minipage}\\
    \begin{minipage}{0.48\textwidth}
        \centering
        \includegraphics[width=\linewidth]{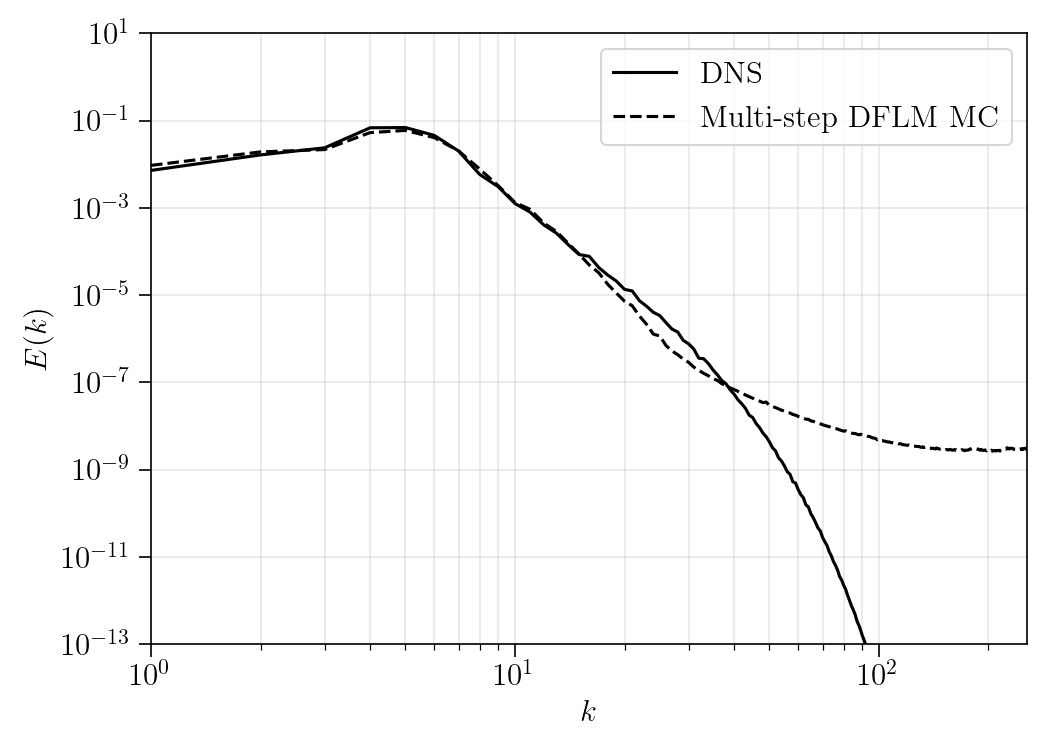}
        (c) Multistep DFLM MC
    \end{minipage}
    \hfill
    \begin{minipage}{0.48\textwidth}
        \centering
        \includegraphics[width=\linewidth]{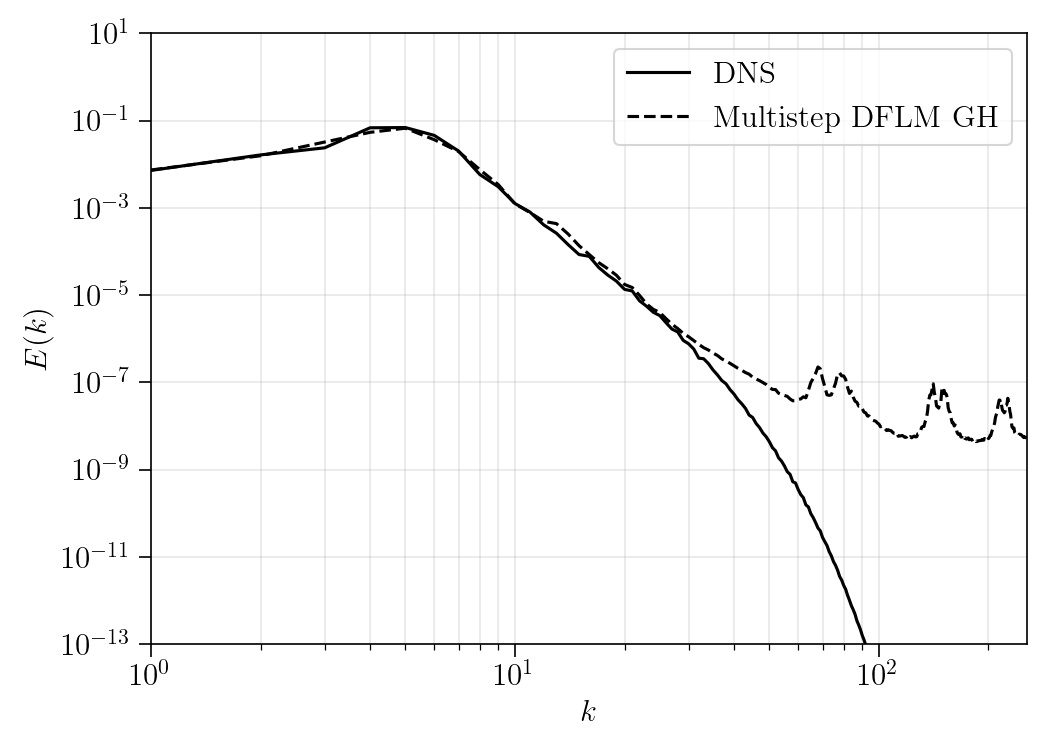}
        (d) Multistep DFLM GH
    \end{minipage}
    \caption{Narrowband-forcing time-averaged energy spectra over $t\in[4.5,5]$. The multistep uses $M=5$ for both DFLM MC and DFLM GH.}
    \label{fig:ex2_avg_spectrum}
\end{figure}

Figure~\ref{fig:ex2_avg_spectrum} compares the time-averaged spectra produced by the Monte-Carlo and Gaussian-approximation targets with one step and with $M=5$ microsteps. In this anisotropic regime, the one-step Monte-Carlo target differs substantially from DNS, while multistep Monte Carlo improves the agreement by resolving part of the within-step deformation through repeated drift evaluations (Section~\ref{sec:DFLMfluid_anisotropy}). The Gaussian-approximation target gives the closest agreement with DNS for both step choices. The residual difference between the multistep Monte-Carlo and Gaussian results suggests that finite-ensemble sampling error is at least as important as the error from neglecting within-step deformation\jh{. Increasing} $M$ addresses the latter only partially, whereas the Gaussian approximation eliminates sampling noise and propagates the transition covariance directly. Thus, in the genuinely turbulent, anisotropic regime, accurate recovery of the macroscopic spectrum depends on resolving the local transition rather than defaulting to an isotropic approximation. The Gaussian target achieves this without requiring a substantially larger walker ensemble.

Section~\ref{sec:multiscale} \jh{provides a physical interpretation of this behavior. Even} for a single microstep, the exact transition covariance already reflects the local strain,
\begin{equation}\label{eq:sigma1_taylor_recall}
\bm{\Sigma}_1 = 2\nu\Delta t \left( \mathbf{I} - \Delta t\,\mathbf{S}_0 \right) + \mathcal{O}((\Delta t)^3), \qquad \mathbf{S}_0 = \tfrac{1}{2}\left( \mathbf{J}_0+\mathbf{J}_0^{\top} \right),
\end{equation}
whereas the one-step Euler--Maruyama covariance in Eq.~\eqref{eq:isotropic_collapse} keeps only the leading, isotropic term. The local strain number
\begin{equation}\label{eq:local_strain_number}
\chi(\bm{x},t) = \Delta t \left\| \mathbf{S}(\bm{x},t) \right\|
\end{equation}
measures \jh{the significance of this correction. As} argued above, the coherent, cascade-reinforced strain of the narrowband example keeps $\chi$ appreciable over the target horizon, whereas it stays small in the broadband example. Because the flow is incompressible, this straining is volume-preserving\jh{, with compression of the distribution along one direction accompanied by stretching along another. Molecular diffusion, though itself isotropic, can therefore produce an anisotropic distribution once transported by the flow.} The exact covariance propagation retains this deformation \jh{whereas} a single Euler--Maruyama step does not.

Multistep Monte-Carlo recovers this deformation only gradually\jh{. The} covariance generated in the first microstep is isotropic, but each subsequent microstep transports and deforms it through the local velocity gradient, so that the covariance after the second microstep already contains
\begin{equation}
(\mathbf{I}-\mathbf{J}_1\delta t) \bm{\Sigma}_1^{\mathrm{EM}} (\mathbf{I}-\mathbf{J}_1\delta t)^{\top},
\end{equation}
which is generally anisotropic. Repeated drift evaluations thus rebuild the missing deformation progressively, and only approximately, whereas the Gaussian approximation captures it exactly within a single microstep.

\section{Discussions and Conclusions}\label{sec:discussion}

This work extends DFLM \cite{DFLMfluid} toward turbulent incompressible flow by directly addressing two limitations identified in that work: reliance on a single-step Euler--Maruyama discretization of the stochastic representation, and the resulting Monte-Carlo sampling error, which \cite{DFLMfluid} identified as an obstacle to applying the method to turbulent flow. Our multi-step discretization, and more effectively our analytic Gaussian approximation of the local stochastic transition, remove this obstacle\jh{. The} Gaussian approximation eliminates the walker-based Monte-Carlo sampling error entirely while also resolving the anisotropic covariance structure that a single isotropic step misses. This addresses sampling error only in evaluating the Bellman target at each collocation point, however\jh{. The} collocation points themselves remain randomly sampled from the spatiotemporal domain to estimate the training loss (Section~\ref{sec:DFLM}), so a separate, outer Monte-Carlo error persists and is not addressed by the present work. The two numerical examples confirm that this matters most precisely where \cite{DFLMfluid} anticipated it would\jh{. In} the genuinely turbulent, strain-organized narrowband-forced flow, the Gaussian approximation gives the best agreement with DNS, whereas in the near-isotropic broadband-forced example the simpler one-step Monte-Carlo target already suffices. Together, the two examples confirm DFLM's efficacy as a multiscale solver for turbulent fluid systems\jh{. It} recovers accurate macroscopic behavior across both the near-isotropic and strongly anisotropic regimes, without resolving the flow on a fine computational mesh.

Interestingly, the improvement gained from multistep Monte-Carlo alone was modest in our experiments, in contrast to \cite{DFLMHomo}, where resolving the target horizon into multiple microsteps was essential for accuracy. The difference reflects what each problem asks the stochastic walkers to resolve. In the homogenization setting of \cite{DFLMHomo}, the governing coefficients carry an explicitly prescribed fine-scale structure, at a known small scale, so accurately sampling that structure requires a correspondingly small microstep. In the present fluid problem, by contrast, neither the forcing nor the velocity field carries such a prescribed sub-target-horizon structure\jh{. The} forcing is constructed at scales chosen directly by us, and the velocity field is itself only ever approximated macroscopically by the network. What limits the one-step transition here is not unresolved fine-scale structure but the anisotropy of the local transition covariance, which the exact Gaussian approximation captures directly regardless of microstepping, and which multistep Monte-Carlo can only approximate indirectly, through the accumulation of many discrete, noisy steps.

This suggests a natural future test problem \jh{ involving fluid flow through a heterogeneous medium,} which would combine the prescribed fine-scale structure that makes multistep essential in \cite{DFLMHomo} with the anisotropic transition covariance addressed here, allowing both mechanisms to be evaluated together.

Beyond this specific test problem, several other directions follow from different aspects of the method developed here. The first concerns the domain\jh{. The} present work considers a periodic domain, and extending the initial and boundary condition treatment of Section~\ref{sec:DFLMfluid} to non-periodic and non-Dirichlet boundaries, particularly perforated domains, is a natural next step. Prior work has applied DFLM to elliptic problems in perforated domains \cite{DFLMperforated}\jh{. Extending} this treatment to the Navier--Stokes setting would allow the method to address fluid systems such as flow through filtration media.

A second direction concerns the modeling assumption at the core of Section~\ref{sec:Gaussian_approximation}\jh{. The} Gaussian approximation is expected to be most accurate when the local transition is well described by its first two moments. Testing the method on fluid systems with more strongly non-Gaussian structure, such as the quasigeostrophic equation \cite{SSPparam} across the low-, mid-, and high-latitude regimes that produce varying degrees of anisotropy, would help delineate the limits of this approximation.

A third, more fundamental direction concerns the scale of macroscopic behavior the method resolves. Section~\ref{sec:DFLMfluid} showed that the target horizon $\Delta t$ governs the size of the neighborhood each stochastic walker explores around its collocation point. A companion theoretical analysis \cite{DFLManalysis} makes this precise at a single collocation point\jh{. Although} the empirical Bellman target is itself unbiased, its sampling variance biases the resulting training loss, which is only asymptotically unbiased as $N_s\to\infty$, with a bias that grows with $\Delta t$ and the spatial gradient of the network but shrinks with the walker count $N_s$\jh{. Consequently,} $\Delta t$ must be bounded below for training to succeed at all, while $N_s$ can be taken as small as desired provided it satisfies the corresponding lower bound set by $\Delta t$. That analysis, however, treats each collocation point in isolation\jh{. It} does not address how the resulting macroscopic scale interacts with the number of collocation points $N_r$ used to cover the domain. The largest macroscopic scale the trained network can resolve is likely also controlled by $N_r$\jh{. The} neighborhoods of nearby collocation points must overlap sufficiently for the network to learn a coherent macroscopic representation across the domain. Making this joint dependence on $\Delta t$, $N_s$, and $N_r$ precise, building on the single-point analysis of \cite{DFLManalysis}, is left to future work.

\section*{Acknowledgments}
JH is suppored by NSF DMS-2607647. YL is supported by NSF DMS-2607646 and ONR MURI N00014-20-1-2595.

\bibliographystyle{siam}
\bibliography{dflm_turbulence}

\end{document}